\documentclass{jfm}
\usepackage{color}
\usepackage{graphicx}
\usepackage{amssymb,amsmath,wasysym}

\usepackage{physics}
\usepackage{tikz,siunitx}
\usepackage{multirow}

\def\bet{\beta}

\graphicspath{{./}{figure/}}

\shorttitle{An isolated logarithmic layer}
\shortauthor{Y. Kwon and J. Jim\'{e}nez}

\title{An isolated logarithmic layer}

\author{Yongseok Kwon
  \corresp{\email{kwon@torroja.dmt.upm.es}}
  \and Javier Jim\'{e}nez}

\affiliation{School of Aeronautics, Universidad Polit\'{e}cnica de Madrid, 28040 Madrid, Spain}

\begin{document}
\maketitle

\begin{abstract}
To isolate the multiscale dynamics of the logarithmic layer of wall-bounded turbulent flows, a novel numerical experiment is conducted in which the mean tangential Reynolds stress is eliminated except in a subregion corresponding to the typical location of the logarithmic layer in channels. Various statistical comparisons against channel flow databases show that, despite some differences, this modified flow system reproduces the kinematics and dynamics of natural logarithmic layers well, even in the absence of a buffer and an outer zone. This supports the {previous} idea that the logarithmic layer has its own autonomous dynamics. In particular, the results suggest that the mean velocity gradient and the wall-parallel scale of the largest eddies are determined by the height of the tallest momentum-transferring motions, implying that the very large-scale motions of wall-bounded flows are not an intrinsic part of logarithmic-layer dynamics. Using a similar set-up, {an isolated} layer with a constant total stress, representing the {logarithmic layer without a driving force}, is simulated and examined.
\end{abstract}

\section{Introduction}

Due to their abundance in scientific and engineering applications, wall-bounded flows have
been one of the key areas in turbulence research. Initially, the focus centred on the
near-wall region due to its direct relation to the generation of skin friction. Over the
last couple of decades, however, the focus has shifted towards the logarithmic layer, partly
because advancement in experimental technique and numerical computing enabled access to flow
databases with a sufficiently resolved logarithmic layer. Yet, a more fundamental reason for
the interest in the logarithmic layer is that it is of great importance in the large-scale
applications of high Reynolds number wall-bounded turbulence (such as large transportation
devices). For instance, the amount of bulk turbulent kinetic energy (TKE) production and
dissipation within the logarithmic layer \citep{Marusic10a,Jimenez18} and the contributions
to the skin friction from the flow structures residing in the logarithmic layer
\citep{deGiovanetti16} both increase with increasing Reynolds number.

One way of investigating the intrinsic dynamics of a particular subregion of the flow is to
isolate it from the rest of the flow. In wall-bounded turbulence, one of the prime examples
is the minimal flow unit of \cite{Jimenez91}. They were able to isolate a quasi-cyclic
sequence of a single set of flow structures in the buffer layer by systematically
restricting the simulation domain, shedding light on its intrinsic dynamics. \cite{Flores10}
extended this approach to the logarithmic layer and identified a series of restricted
minimal domains in which turbulence characteristics are well-replicated up to a wall-normal
distance proportional to the spanwise domain size. They used them to study the
characteristics of a hierarchy of the minimal logarithmic layer structures without an
influence from the large-scale outer layer motions. Later, \cite{Hwang15} combined this
method with an overdamped large eddy simulation (LES) \citep[use of intentionally elevated
value of eddy viscosity to damp out the small-scale motions; see][]{Hwang10} to isolate flow
structures of a given step in the hierarchy. Based on this experiment, he concluded that the
flow structures at each hierarchy can sustain themselves. However, there is still a question
of whether overdamping simply filters out the small-scales (without affecting the
large-scales) or modifies the dynamics of the whole flow by effectively reducing the
Reynolds number \citep{Feldmann18}.

While the above examples attempt to isolate structures of certain sizes or locations, the
study of statistically stationary homogeneous shear turbulence (SSHST) takes the different
approach of isolating a particular element of the logarithmic layer, namely the shear.
Unlike experimental homogeneous shear flows, where the size of the structures grows indefinitely,
an statistically stationary state is achieved numerically by using a limited spanwise flow
domain \citep[e.g.][]{Pumir96,Sekimoto16}. This flow setup lacks near-wall dynamics, and is
hence suitable for investigating the isolated effect of the mean shear on the flow dynamics.
\cite{Dong17} conducted an extensive study of the coherent structures in SSHST, and
concluded that its  structures are essentially  symmetrised
and unconstrained (by the wall) versions of the structures in turbulent channels, suggesting
that the shear is the main ingredient of the coherent structure dynamics in the logarithmic
layer. However, SSHST is still not equivalent to the logarithmic layer, because it cannot
replicate the wall-normal dependence of characteristic length scale, or the inhomogeneity
along the wall-normal direction.

In this regard, a closer reproduction of the logarithmic layer is the numerical experiment
by \cite{Mizuno13} where the buffer layer (as well as the wall itself) was removed and
substituted by an off-wall boundary condition. They introduced the scale variation along the
wall-normal direction by using a rescaled interior plane as the off-wall boundary (without
the rescaling, the resulting flow was very similar to SSHST). Their numerical experiment
reproduced many characteristics of the natural logarithmic layer, although not perfectly.
For example, a spurious `buffer layer' formed near the off-wall boundary due to the
formation of small-scale vortices caused by the incoherence between the off-wall boundary
and the adjacent flow. Alternatively, \cite{Lozano-Duran19b} achieved the same objective by
utilizing slip and permeable boundary conditions. This experiment reproduces the outer layer dynamics of the no-slip channel well but only does so above some adaptation height, which is of the order of the slip length applied for the boundary conditions. In combination with that, \cite{Bae19}
used a minimal spanwise domain to remove the large-scale outer layer structures to isolate
the logarithmic layer.

All the aforementioned studies were successful at replicating or isolating certain features
of the logarithmic layer, but also had some drawbacks which made them incompatible with the
natural flow. In the previous attempts to isolate the logarithmic layer of turbulent channel
flows, there have been numerous strategies for removing the buffer layer dynamics. However,
to our best knowledge, the removal of the outer layer large-scale structures has relied
almost exclusively on the use of a minimal spanwise domain. In this work, an alternative
strategy is employed to remove the large-scale outer motions by modifying the driving force
of the flow. It has an advantage that the large-scale structures are removed without
artificially saturating their wall-parallel growth. This method is somewhat similar to the
method of \cite{Jimenez99} where they introduced an explicit damping term in the evolution
equations of the flow above the buffer layer to remove the outer layer motions. The
resulting flow had a laminar outer flow while the undisturbed part of the flow displayed
similar behaviour to the near-wall turbulence, although some of flow statistics were
altered. However, in the present work, the evolution equations of the flow are not modified
except by the body force. Therefore, this investigation aims to isolate the logarithmic
layer of turbulent channel flows with a minimal disturbance to its essential dynamics.

{As in most of the examples just mentioned, the system that we analyse here is only an approximation to the canonical logarithmic layer. As in those cases, it is best understood as an example of the `thought experiments' that have been a mainstay of physics for a long time. It is intended to represent the logarithmic layer in the same sense that point masses are often used to represent planets. In all these cases, it is equally important to recognise which features are retained by the approximation and which ones are not. We will see below that some properties that could be expected to depend on the near-wall region (e.g. the self-similar hierarchy of attached eddies) are well reproduced by our approximation, even if that region is missing from our model. On the other hand, properties linked to the outer flow are not well reproduced, and this will be used to explain the origin of some of the features observed in true logarithmic layers. It is also important to emphasise that the logarithmic layer requires a theory that cannot simply be provided by increasing the Reynolds number of the simulations. In intermediate asymptotic ranges, such as the logarithmic layer or the inertial range of isotropic turbulence, the theory for the self-similar regime requires being able to separate its dynamics from the details of its interactions with the inner and outer limits \citep{Barenblatt96}. Nonetheless, those details are often important in themselves. For example, the interaction of the logarithmic layer with the near-wall layer is key to formulating correct boundary conditions for large-eddy simulations \citep{Jimenez00}, while the interaction with the outer flow is required to understand why and how properties like the turbulence intensities depend on the Reynolds number \citep{Hutchins07b}.}

The organization of this paper is the following. \S 2 outlines the details of the numerical experiments, and \S 3 assesses the quality of the isolated logarithmic layer. Finally, the major findings of this paper are {discussed in \S 4 with conclusions in \S 5}.

\section{Numerical experiment}

\begin{table}
	\centering
	\begin{tabular}{cccccccccc}
		\hline \noalign{\smallskip}
		Case & Type & $Re_\tau$ & $L_x/h$ & $L_z/h$  & $\Delta x^+$ & $\Delta z^+$ & $\Delta y^+$ & $tU_\tau/h$ & Line style \\
		\hline \noalign{\smallskip}
		LB & LES & 2002 & $2\pi$ & $\pi$ & 37.0 & 37.0 & 2.36 - 13.0 & 24.2 & {\color{black}\tikz[baseline]{\draw[thick,dashdotted] (0,.5ex)--++(.5,0) ;}}\\
		LW & LES & 1998 & $2\pi$ & $\pi$ & 36.9 & 36.9 & 2.35 - 13.0 & 27.0 & {\color{black}\tikz[baseline]{\draw[thick] (0,.5ex)--++(.5,0) ;}}\\
		LN & LES & 2000 & $2\pi$ & $\pi$ & 36.9 & 36.9 & 2.36 - 13.0 & 30.1 & \tikz[baseline]{\draw[thick,dashed] (0,.5ex)--++(.5,0) ;}\\
		LWc & LES & $\infty$ & $2\pi$ & $\pi$ & 36.9 & 36.9 & 1.57 - 8.65 & 29.5 & {\color{black}\tikz[baseline]{\draw[thick] (0,.5ex)--++(.5,0) node [midway, draw=black, fill=white, shape=circle, minimum size=3.0pt, inner sep=0pt] {} ;}}\\
		HJ06 & DNS & 2003 & $8\pi$ & $3\pi$ & \multicolumn{4}{c}{\cite{Hoyas06}} & {\color{black}\tikz[baseline]{\draw[thick,dotted] (0,.5ex)--++(.5,0) ;}}\\
		\hline
	\end{tabular}
\caption{Simulation parameters for the numerical experiments, and for the reference DNS
database. $\Delta$ represents the grid spacing in each direction. The grid
spacings in the wall-parallel directions are calculated after dropping 1/3 of the high
wavenumber modes for de-aliasing. The LES cases include the base case (LB), the main
experiment with wider logarithmic layer (LW), the supplementary experiment with narrower
logarithmic layer (LN) and the experiment with a constant stress profile (LWc). For LW, LWc
and LN, $U_\tau$ is computed by extrapolating the total shear stress to the wall. For LWc,
$h$ is the wall-normal simulation domain. The next-to-last column shows the total time over
which the statistics are gathered, in terms of the large-eddy turnover time ($h/U_\tau$).}
	\label{tb:parameters}
\end{table}

For this investigation, turbulent flow between two parallel plates, separated by the
distance $2h$, is simulated at a nominal $Re_\tau=hU_\tau/\nu=2000$, where $\nu$ is the
kinematic viscosity of the fluid and $U_\tau$ is the friction velocity. Periodic boundary
conditions are used along the wall-parallel directions and no-slip and impermeable boundary
conditions are applied at both walls. Throughout the paper, the streamwise, wall-normal and
spanwise coordinates are denoted by $x$, $y$ and $z$, respectively, and the corresponding
velocity components by $U$, $V$ and $W$. The $y$-dependent ensemble-averaged quantities are
represented by an overline, while lower-case velocity variables indicate fluctuations with
respect to this average (e.g. $U=\overline{U}+u$). A `+' superscript indicates normalisation
by the viscous scale $\nu/U_\tau$ for length, and by $U_\tau$ for velocity. The domain
length in $x$ and $z$ are $L_x=2\pi h$ and $L_z=\pi h$, respectively, to make sure that the
entire flow domain is not minimal in the wall-parallel directions \citep{Flores10}. The flow
is simulated via LES with static Smagorinsky sub-grid scale (SGS) model
\citep{Smagorinsky63}. The Smagorinsky constant is chosen to be $C_s=0.1$, and the
statistics of the LES compare well with a direct numerical simulation (DNS) database at the
same Reynolds number. The computational algorithm and the numerical code are taken from
those employed in \cite{Vela-Martin19}, but adapted for LES.
{The code solves the wall-normal vorticity and the Laplacian of $v$ formulation of the Navier-Stokes equations \citep{Kim87}. Along the wall-parallel directions, the equations are projected onto Fourier basis functions along a uniform mesh. A non-uniform mesh is used in the wall-normal direction to account for the inhomogeneity in that direction, and the wall-normal gradients are computed by using seven-point compact finite differences with spectral-like resolution \citep{Lele92}. For temporal integration, a low-storage semi-implicit third-order Runge-Kutta scheme is used \citep{Spalart91}.}

Since the purpose of the current
experiment is to isolate the dynamics of the logarithmic layer, it has to adequately
resolve the energy-containing motions in that region. For this purpose, a DNS database of
channel flow at the comparable $Re_\tau$ \citep[][hereafter referred to as HJ06]{Hoyas06} is
examined as a benchmark, and it is found that more than $95\%$ of the total turbulent
kinetic energy, $k=(\overline{u^2}+\overline{v^2}+\overline{w^2})/2$ is contained
within motions whose streamwise and spanwise wavelengths are larger than 74 viscous
length units for $y^+>100$. Therefore, the nominal grid spacings in $x$ and $z$ are chosen
to be $\Delta x^+=\Delta z^+\simeq37$, after de-aliasing. In the wall-normal direction, the
grid is defined by a hyperbolic tangent stretch function such that the $n$-th grid location
is given by $y_n/h=\tanh[3(n-1)/511-3/2]/\tanh(3/2)+1$ for $n=1,2,...,512$, between the
lower wall at $y=0$ and the upper one at $y=2h$. The parameters of the simulations are
summarized in table \ref{tb:parameters}.
{Here, $Re_\tau$ for LW, LWc and LN are given based on $U_\tau$ from the extrapolated total shear stress at the wall to highlight the agreement of the total stress profile within linear-stress layer. However, it does not carry the usual meaning of the `Reynolds number' for the canonical channel flows because the scale separation and the wall-normal gradient of the total shear stress become independent parameters for our isolated layers (for the canonical channel flows, they are both related by $Re_\tau$).}

\begin{figure}
\centering
\includegraphics[width=0.8\linewidth]{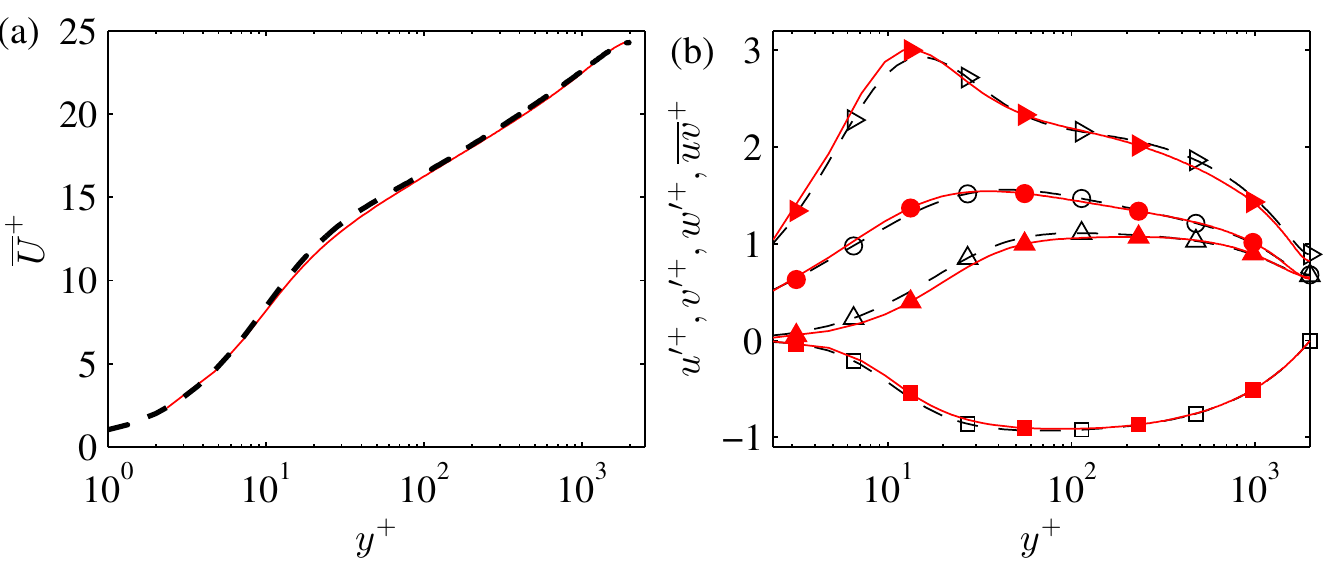}
\caption{(a) Mean streamwise velocity. Solid line: LB, Dashed line: HJ06. (b) Second-order velocity statistics. $\rhd$: $u'^+$, $\triangle$: $v'^+$, $\Circle$: $w'^+$, $\Square$: $\overline{uv}^+$. Solid lines with closed symbols: LB, Dashed lines with open symbols: HJ06.}
\label{fig:stat_base}
\end{figure}

In order to validate this numerical methodology, the base LES (case LB) is first conducted
and compared against HJ06. For this case, a van Driest damping function of the form
$D(y^+) = [1-\exp(-y^+/26)]^2$ is used on the Smagorinsky eddy viscosity to enforce the zero
SGS stress conditions at the wall. Figure \ref{fig:stat_base} shows the profiles of the mean
streamwise velocity $\overline{U}$, and of the second-order velocity statistics of LB. The
primed velocity variables indicate the root mean square (RMS) value. A good agreement is
observed between the statistics of LB and HJ06. Although not shown here for brevity, the
one-dimensional (1D) velocity spectra also show a good agreement. Throughout the paper,
further statistical comparisons will be made where appropriate to demonstrate that our LES
code reproduces well the logarithmic layer of turbulent channel flows.

Several methods were tested to isolate the logarithmic layer by removing the turbulent
fluctuations outside it. The initial strategy was to employ an elevated value of $C_s$
outside the logarithmic layer \citep[i.e. overdamped LES; see][]{Hwang10} and its effect on
the flow was investigated by varying the value of $C_s$ in the usual buffer layer
($y^+<70$). However, with increasing $C_s$ near the wall, it is observed that the spectral
signature of the near-wall cycle gradually moves outwards instead of being eliminated at a
fixed location {(see Appendix \ref{appA})}.

Hence, instead of damping previously-created turbulent fluctuations, an alternative approach
is sought where the necessity of `active' turbulent fluctuations is eliminated outside the
logarithmic layer. This is achieved by setting a prescribed total mean shear stress (sum of
viscous, Reynolds and SGS stresses) profile which drops to zero outside the \textit{nominal}
logarithmic layer. In practice, it is done by imposing a modified profile of the body force.
This method is found to be effective at eliminating the buffer layer, and is chosen as our
preferred method for isolating the logarithmic layer. 
{The method of modifying the stress profile also means that the elimination of the outer layer dynamics can be achieved without relying on the restricted flow domain and hence allows us to investigate its effects on the large-scale structures, which is not possible in the case of the minimal logarithmic layer experiments where the large-scales are, by construction, truncated.}
In fact, the idea is not new. For
example, \cite{Tuerke13} simulated turbulent channel flows with a prescribed mean velocity
profile to study its effects on the dynamics of energy containing eddies, and
\cite{Borrell15} applied and extra body force to model the effects of roughness near the
wall. It is also known that a modified body force can lead to laminarization of the flow at
transitional Reynolds numbers \citep{He16,Kuhnen18}. {\cite{Russo16} investigated the linear response of the mean streamwise velocity of turbulent channel flows to a body force, albeit at low Reynolds number.} However, to our best knowledge, it has
not been used for the purpose of isolating a particular subregion of the flow.

\begin{table}
	\centering
	\begin{tabular}{lcccccccccc}
		\hline \noalign{\smallskip}
		\multirow{2}{*}{Case} & \multirow{2}{*}{Equation} & \multicolumn{4}{c}{Body force} && \multicolumn{2}{c}{Linear layer} && \multirow{2}{*}{$\delta_a/h$} \\
		\cline{3-6}\cline{8-9}
		\rule{0pt}{2.5ex} 
		 & & $y_{l}/h$ & $\bet_{l}$ & $y_{u}/h$ & $\bet_{u}$ && $y_{bot}/h$  &  $y_{top}/h$ &&  \\
		\hline \noalign{\smallskip}
		LW & (\ref{eq:stress}) & 0.025 & 120 & 0.35 & 20 && 0.045 & 0.235 && $0.449$ \\
		LWc & (\ref{eq:stress_constant}) & 0.025 & 120 & 0.35 & 20 && 0.045 & 0.235 && $0.466$ \\
		LN & (\ref{eq:stress}) & 0.025 & 120 & 0.15 & 60 && 0.045 & 0.11 && $0.188$ \\
		\hline
	\end{tabular}
\caption{Parameters for the prescribed stress profiles. The linear-stress layer is the
region where the deviation from the natural stress profile ($\overline{\tau}_{xy}^+=1-y/h$
for LW and LN, and $\overline{\tau}_{xy}^+=1$ for LWc) is below $1\%$. $\delta_a$ represents
the height of the active stress region where $\overline{\tau}_{xy}^+>0.01$.}
	\label{tb:stress_profile}
\end{table}

\begin{figure}
\centering
\includegraphics[width=0.5\linewidth]{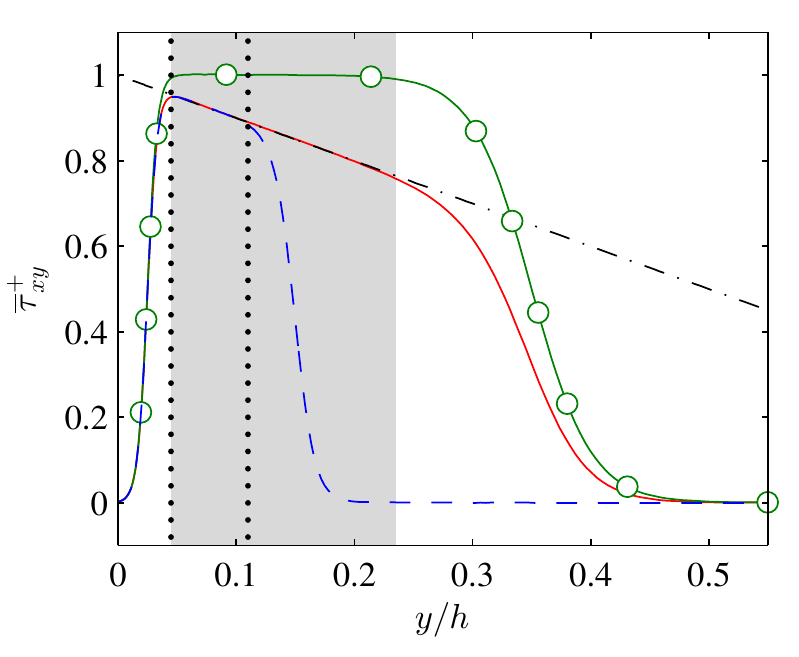}
\caption{Profiles of the mean shear stress. The lines are as indicated in table
\ref{tb:parameters}, except for the dashed-dotted line, $\overline{\tau}_{xy}^+=1-y/h$ (LW,
LWc and LN are presented by red, green and blue colors, respectively). The linear-stress
layer for LW and LWc is indicated by the grey shaded area and that for LN is indicated by
the vertical dotted lines.}
\label{fig:stress}
\end{figure}

Two numerical experiments are performed with different prescribed mean stress profiles, as
shown below and in figure \ref{fig:stress},
\begin{equation}
\overline{\tau}_{xy} = \frac{U_\tau^2}{4}(1-y/h)\left(1+\tanh[\bet_l(y/h-y_l/h)]\right)(1-\tanh[\bet_u(y/h-y_u/h)]) .
\label{eq:stress}
\end{equation}
This equation is defined for $0 \leq y \leq h$, but the prescribed stress profile is
extended to the opposite side of the channel using symmetry. The parameters $y_l$ and
$\bet_l$ control the location and width of the region where the stress profile decays from
its natural value to zero between the nominal logarithmic layer and the wall. Likewise,
$y_u$ and $\bet_u$ control the location and width of the region where the stress profile
decays smoothly to zero above the nominal logarithmic layer. The parameters for the stress
profiles for the main experiment with isolated logarithmic layer (case LW) and for the
supplementary experiment with narrower log layer (case LN) are given in table
\ref{tb:stress_profile}. The values of $y_l$, $\bet_l$, $y_u$ and $\bet_u$ are set
empirically.

In addition to  cases LW and LN, another experiment is performed in which the prescribed
stress profile is constant within the isolated layer, intended to mimic the total stress
profile in the logarithmic layer as the {streamwise pressure gradient vanishes}. For this
experiment, referred to as LWc, the prescribed stress profile is 
\begin{equation}
\overline{\tau}_{xy} = \frac{U_\tau^2}{4}\left(1+\tanh[\bet_l(y/h-y_l/h)]\right)(1-\tanh[\bet_u(y/h-y_u/h)]) , 
\label{eq:stress_constant}
\end{equation}
which only differs from (\ref{eq:stress}) by the missing $(1-y/h)$ factor. The parameters
$y_l$, $\bet_l$, $y_u$ and $\bet_u$ are kept as in LW, to study the effect of changing the
stress profile independently from other factors. In conventional channel flows, the channel
centreline provides a natural symmetry plane for the total stress profile (i.e.
$\overline{\tau}_{xy}=0$). This does not apply to LWc, where the extrapolated location of zero mean shear
stress would be infinitely far from the wall. Therefore, the upper wall of LWc is replaced
by a free-slip impermeable boundary at $y=h$ ($\partial u/\partial y=\partial w/\partial
y=0$ and $v=0$). Although physically not required, fine grid spacing is needed near the
free-slip wall to numerically enforce the boundary conditions (in particular, to numerically
resolve exponential functions with large exponents for high wavenumbers). Hence, a different
wall-normal grid is used for LWc, such that the $n$-th grid location is given by
$y_n/h=\tanh[3(n-1)/383-3/2]/(2\tanh(3/2))+1/2$ for $n=1,2,...,384$ for $0\leq y_n\leq h$.
This grid is designed such that the wall-normal spacing is kept similar to that used for LB,
LW and LN within the domain of interest ($0.05\lesssim y/h\lesssim 0.2$). It should be noted
that $h$ simply means the wall-normal domain size for LWc, not the channel half height.

We define the `linear-stress', or `active', layer to be the region in which the prescribed
shear-stress profile deviates by less than $1\%$ from the natural stress profile in channels
with unmodified body force (i.e. $\overline{\tau}_{xy}^+=1-y/h$ for LW and LN, and
$\overline{\tau}_{xy}^+=1$ for LWc). While eddies within this layer could be expected to be
`most natural',  taller eddies have to exist up to the level at which some tangential
stress has to be carried by the flow. We therefore also introduce a length scale,
$\delta_a$, intended to be indicative of the height of the largest momentum-transferring
eddies, defined as the point at which $\overline{\tau}_{xy}^+=0.01$ (approximately $1\%$
of the wall shear stress in the natural channel). The limits $y_{bot}<y<y_{top}$ of the
linear layer, and $\delta_a$ of the active one, are given in table \ref{tb:stress_profile}.
Note that, although $y_{top}$ and $\delta_a$ are related, they are independent parameters,
whose ratio can be changed by modifying the stress profile. Both will be used below to scale
different quantities, but, since $y_{top}/\delta_a\approx 0.5-0.6$ in all our experiments,
it is impossible to say which of the two, if any, is the most physically relevant scale. On the other
hand, it follows from the definition of $\delta_a$ that we can tentatively assign $\delta_a=h$ in
unmodified channels.
 
Within the linear-stress layer, the flow experiences a body force equivalent to the one
generated by the mean pressure gradient in a canonical channel, and most of the stress is
carried by the Reynolds stress. For example, the flow in LN and LW  has to produce the same
mean momentum flux as in a natural channel at $Re_\tau=2000$, and could therefore be
expected to have the same dynamics as the logarithmic layer in such a channel. For LWc, no
driving body force is present in the linear region, and the flow is maintained by the
localised body forces applied above and below the linear-stress layer. For the actual
numerical computation, the stress profile is enforced by replacing the mean pressure
gradient in the mean streamwise momentum equation with $-\dd\overline{\tau}_{xy}/\dd y$. No
van Driest damping is used, since the buffer layer is outside the domain of validity of the
experiments, and there is no need to reproduce its behaviour. Figure \ref{fig:stress} shows
the actual stress profiles for LW, LWc and LN, which follow the prescribed profiles well.
For all the numerical experiments in which the wall shear stress is intentionally modified,
the effective $U_\tau$ is estimated by extrapolating to the wall the stress in the
linear-stress layer. This is the velocity scale used for normalization in table
\ref{tb:parameters}.

\begin{figure}
\centering
\includegraphics[width=0.8\linewidth]{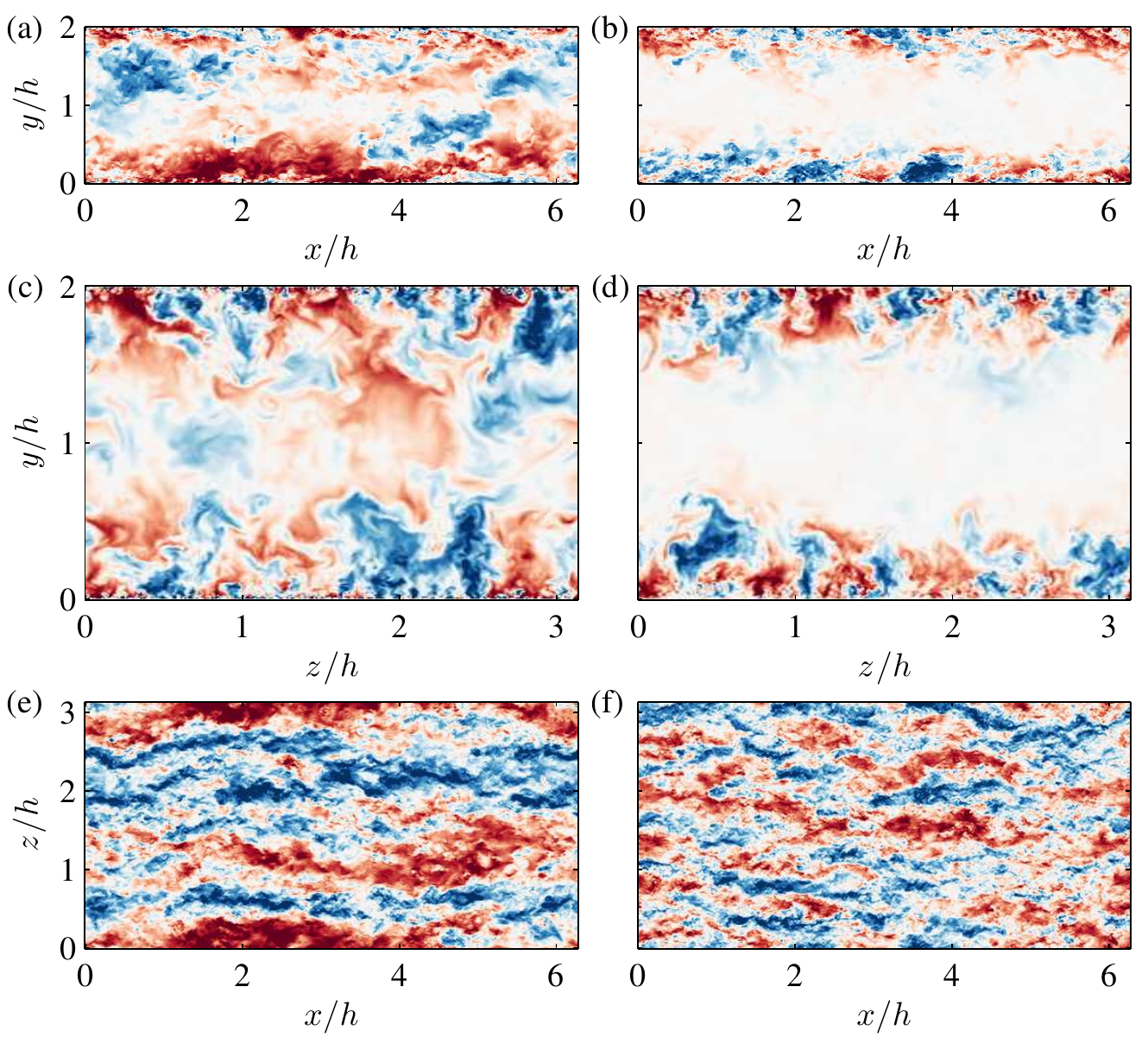}
\caption{Instantaneous fields of $u$ for (a,c,e) LB; (b,d,f) LW. The wall-parallel planes in (e,f) are at $y/h=0.15$. The colors range from $u=-4U_\tau$ (dark blue) to $4U_\tau$ (dark red).}
\label{fig:inst_fields}
\end{figure}

As a preliminary result and a qualitative comparison, instantaneous snapshots of the  field of $u$
for LB and LW are shown in figure \ref{fig:inst_fields}. They demonstrate that the turbulent
fluctuations in the centre of the channel are eliminated in LW. This is also true in the buffer layer,
although they are too small to be observed visually. Within the isolated layer, the
turbulent structures are qualitatively similar in both cases, although it is noteworthy that
streaks whose streamwise length is comparable to the streamwise domain are observed in LB
but not in LW. This difference will be further investigated by the spectral analysis in \S 3.2.

\section{Assessment of the isolated logarithmic layer}

\subsection{One-point statistics}

\begin{figure}
\centering
\includegraphics[width=0.8\linewidth]{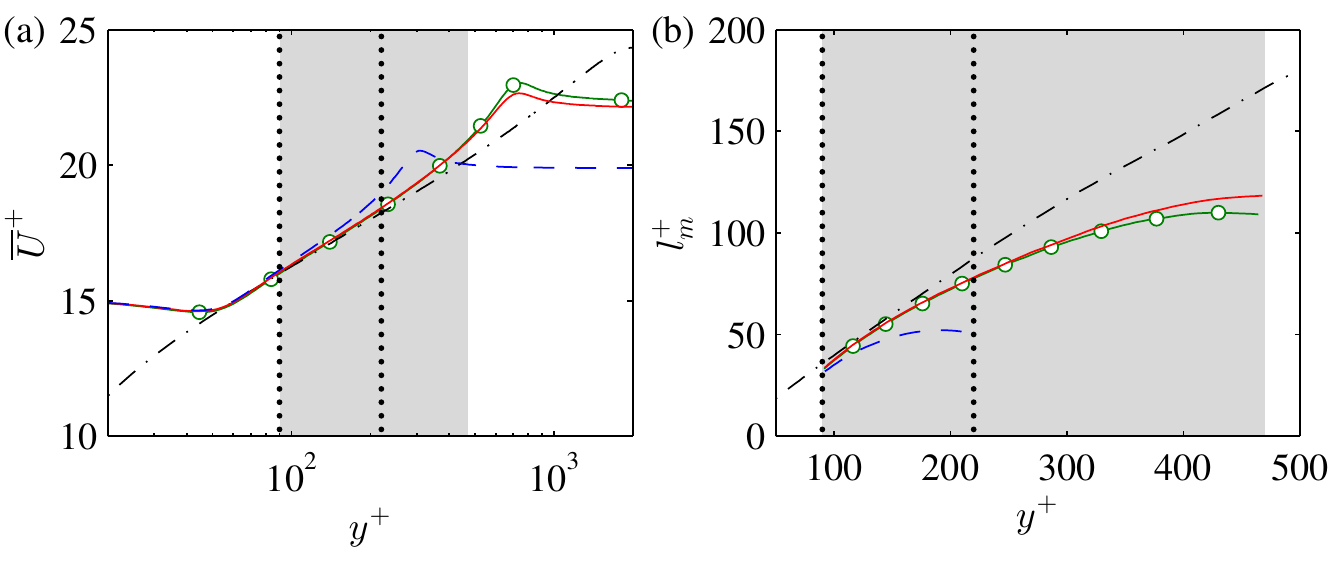}
\caption{(a) Mean streamwise velocity (shifted by $15U_\tau$ for LW, LWc and LN) (b) Mixing length. For LW, LWc and LN, only the linear shear stress region is shown. The lines are as indicated in table \ref{tb:parameters} (LB, LW, LWc and LN are presented by black, red, green and blue colors, respectively). The linear-stress layer for LW and LWc is indicated by gray shaded area and that for LN is indicated by vertical dotted lines.}
\label{fig:stat_U}
\end{figure}

In this section, we examine whether the flow in an isolated linear-stress layer can replicate the
characteristics of the natural logarithmic layer, by comparing the statistics of the
truncated cases, LN and LW, with those of the full channel LB. In addition to that, we
examine the effects of changing the stress profile by comparing case LW with LWc, which is
intended to represent the limiting case of channel flow {without the driving force}. Figure
\ref{fig:stat_U}(a) shows the profiles of the mean streamwise velocity. Note that, because
the walls are outside the domain of validity of the three isolated cases, the no-slip
condition does not provide an absolute velocity reference, and a Galilean offset of the
profile is required in general \citep{Mizuno13}. In fact, it is clear from the figure
that the mean velocity of the truncated layers vanishes below $y^+\approx 60$ (and actually
becomes slightly negative). Much of the effect of the no-slip condition is taken over by the
dragging effect of the body force, and the profiles for LW, LWc and LN need to be shifted by
$15U_\tau$ to be comparable to the canonical logarithmic layer. The agreement of the mean velocity after this
shift is fair within the active layer, but the velocity gradient gets steeper as the
linear-stress layer gets narrower. This is further examined in figure \ref{fig:stat_U}(b),
which tests the  mixing length, $l_m=U_\tau/S$, where
$S=\dd \overline{U}/\dd y$. For a logarithmic mean velocity, $l_m^+ (y^+)$ is a linear
function whose slope is the K\'{a}rm\'{a}n constant, but the mixing-length profile of LW,
LWc and LN is not linear, even within the active layer.
{By comparing wall-bounded flows with different geometries (with an exception of Ekman layers), \cite{Johnstone10} and \cite{Luchini17} found that in the logarithmic and outer layers, the negative streamwise pressure gradient induces a positive shift in the mean streamwise velocity, and vice versa. In our experiments, the mean streamwise velocity of LWc is higher than that of LW, which seems contradictory to those previous results. However, a direct comparison is not possible here because the positive shift in $\overline{U}$ profile of LWc with respect to LW is caused by the difference in the wall-normal profile of the body force, rather than by the geometry or pressure gradient. Although the total integrated force must sum to zero in both LW and LWc, the magnitudes of integrated positive and negative forces (i.e. the difference between the minimum and maximum $\overline{\tau}_{xy}$) are about $5\%$ larger in LWc, which results in the greater amount of mean shear and the positive shift in $\overline{U}$.}

\begin{figure}
\centering
\includegraphics[width=0.8\linewidth]{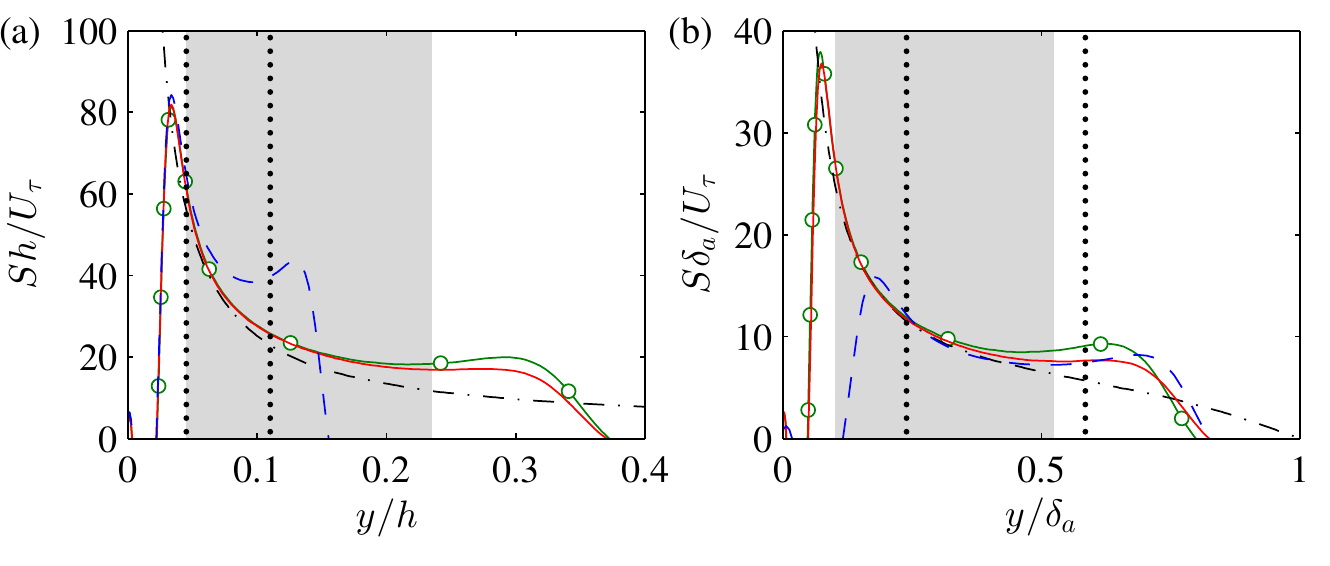}
\caption{The mean shear profile scaled by (a) the channel half height; (b) the width of active stress layer. The lines are as indicated in table \ref{tb:parameters} (LB, LW, LWc and LN are presented by black, red, green and blue colors, respectively). The linear-stress layer for LW and LWc is indicated by gray shaded area and that for LN is indicated by vertical dotted lines.}
\label{fig:stat_S}
\end{figure}

{The difference in the mean velocity} is further investigated by comparing the mean shear profiles of  the four LES
cases. Figure \ref{fig:stat_S}(a) shows that they collapse poorly when $S$ is normalised by
$U_\tau$ and $h$. However, in our experiments, $h$ does not convey the usual meaning of an
outer length scale, because the eddies of height $h$ are purposely suppressed. Instead, we
propose that the alternative length scale, $\delta_a$, is more relevant to the physics of
the flow, since it represents the height of the tallest momentum-transferring eddies. Figure
\ref{fig:stat_S}(b) shows that the profiles collapse well within the linear-stress layer
when both $S$ and $y$ are normalised with $\delta_a$, at least up to $y\simeq 0.4\delta_a$.
It is particularly interesting that the profile of the mean shear scales with
$\delta_a$ even when the profile of total shear stress (which also represents the driving
force) changes within the active layer, such as between LW and LWc. This suggests that the
value of shear within the logarithmic layer is associated with, or possibly decided by,
the size of the largest active eddies in the flow, and that this size is controlled by
$\delta_a$. Moreover, the fact that the profiles agree within the active layer, when
properly scaled, suggests that the truncated flows contain a self-similar eddy hierarchy, as
in the natural logarithmic layer, although the range of sizes within the hierarchy may differ.

\begin{figure}
\centering
\includegraphics[width=0.8\linewidth]{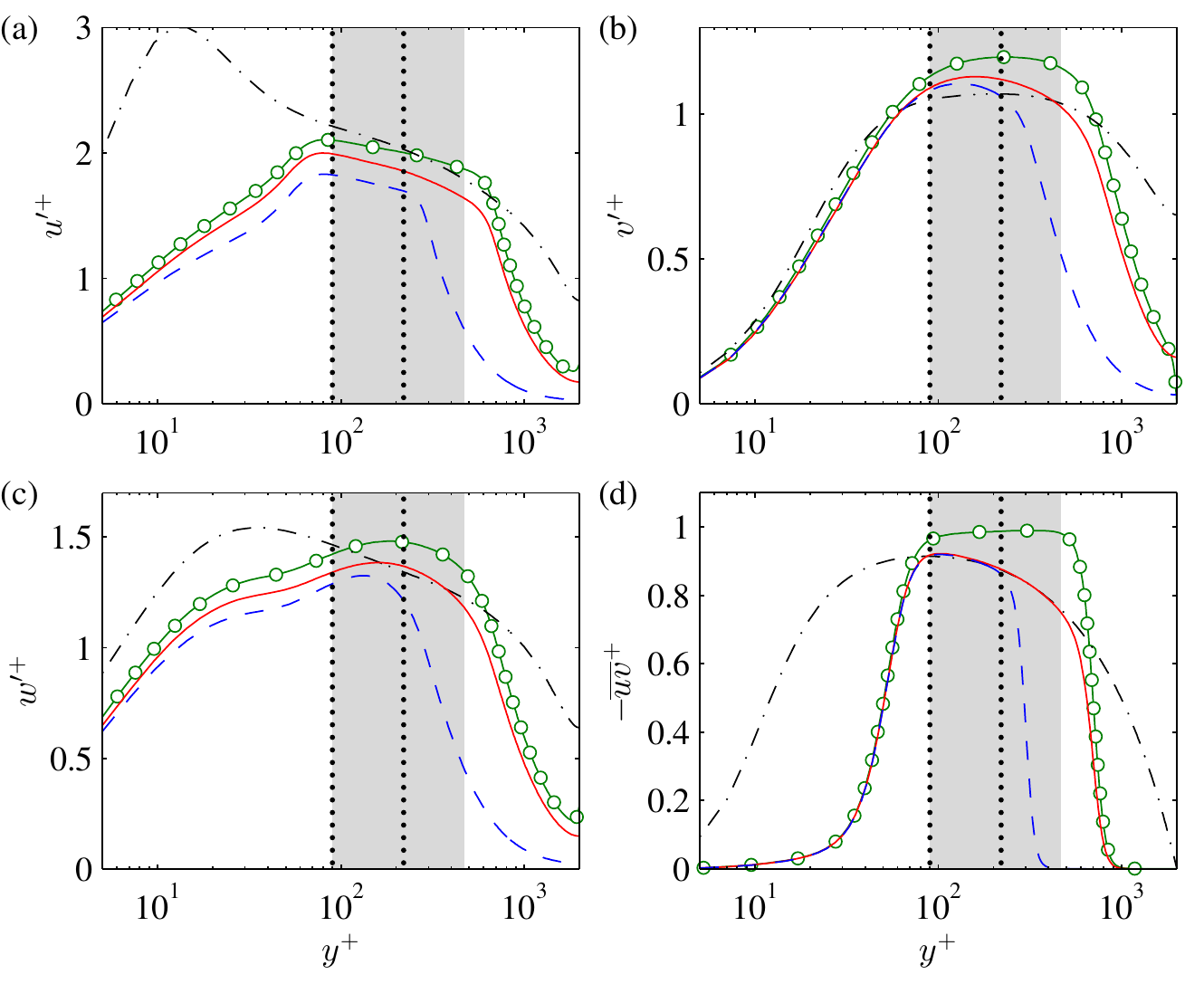}
\caption{Profiles of (a) $u'^+$ (b) $v'^+$ (c) $w'^+$ (d) $-\overline{uv}^+$. The lines are as indicated in table \ref{tb:parameters} (LB, LW, LWc and LN are presented by black, red, green and blue colors, respectively). The linear-stress layer for LW and LWc is indicated by gray shaded area and that for LN is indicated by vertical dotted lines.} 
\label{fig:stat_rms}
\end{figure}

Figure \ref{fig:stat_rms} compares the Reynolds stress profiles of the four LES cases. All
the stresses decay for $y\gtrsim 2\delta_a$, and figure \ref{fig:stat_rms}(d) shows that the
shear stress agrees well within the active layer for LB, LW and LN, as expected from the
design of the experiment. An important observation is the absence of a buffer-layer $u'$
peak in LW, LWc and LN, which suggests that the buffer-layer dynamics has been suppressed.
There are some residual velocity fluctuations below the linear-stress layer, but they are
not involved in the net momentum transfer or in TKE production, because they only carry a
negligible fraction of the tangential Reynolds stress (i.e. they are inactive, see figure
\ref{fig:stat_rms}d). The shape of the $u'$ profiles within the linear-stress layer is
similar for LB, LW and LN, but their amplitude decreases as the width of the active layer
does. The same decreasing trend is observed for $w'$ when comparing LW and LN, and we will
argue below that both trends are due to the attenuation of the large-scale fluctuations by
the restricted height of the active layer. Here, the effects of changing the height of the
active layer can be solely attributed to the change in the scale separation within the eddy
hierarchy, because LB, LW and LN share the same mean shear stress within the linear-stress
layer. In contrast, the value of $v'$ is slightly higher for LW and LN than for LB. The
exact reason for this is not clear, but the likeliest explanation is that an elevated $v'$
is required to compensate for the missing tangential  Reynolds stress that used to be contributed by the 
large-scale $u$-eddies  that would otherwise have originated above the active region
(see figure \ref{fig:inst_fields}). In the outer part of the flow, where
$y\sim\mathcal{O}(\delta_a)$, the profiles of the truncated simulations collapse well when
$y$ is scaled with $\delta_a$ (not shown), indicating that they have similar outer layer
dynamics.

\begin{figure}
\centering
\includegraphics[width=0.8\linewidth]{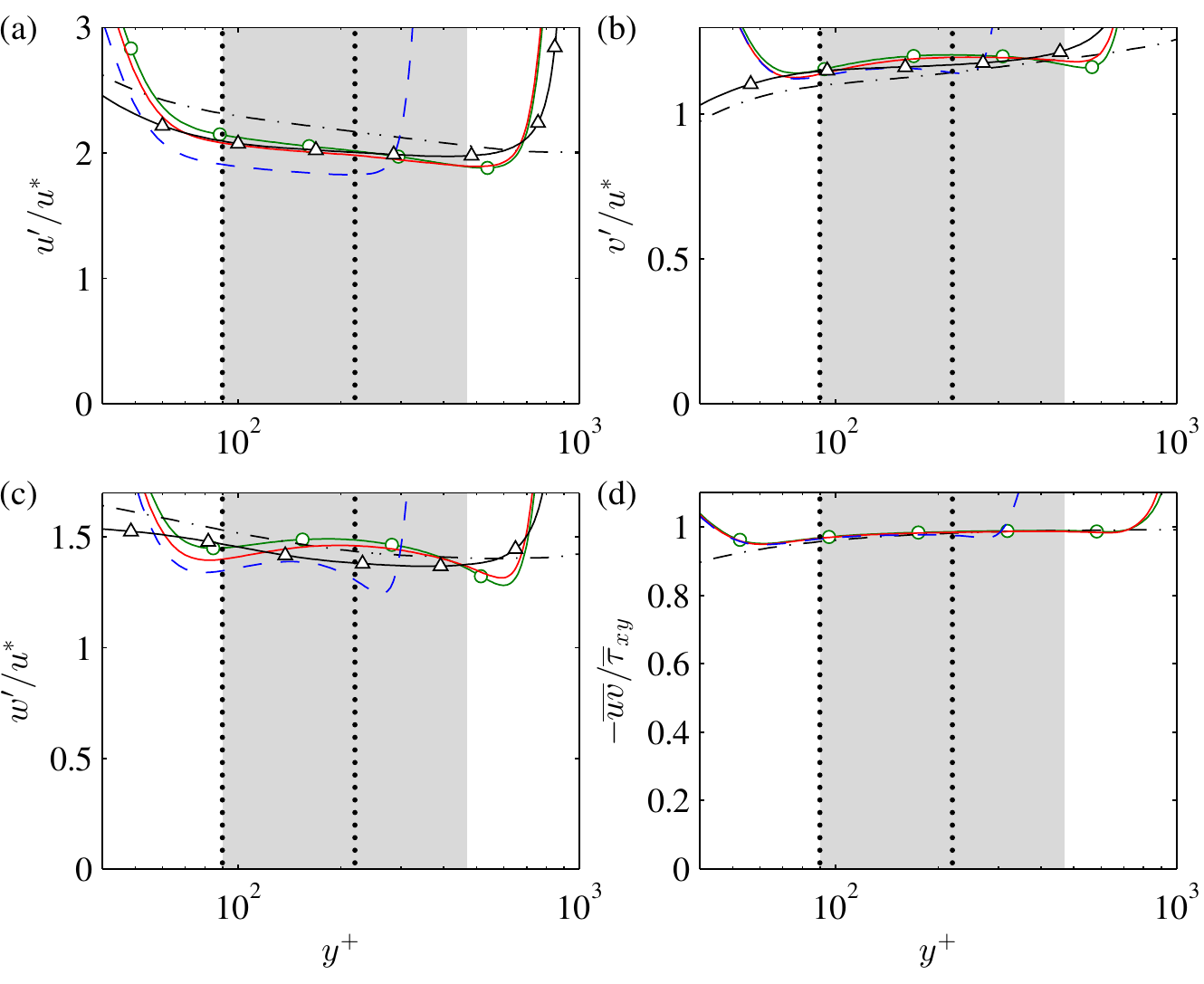}
\caption{Profiles of (a) $u'$ (b) $v'$ (c) $w'$ (d) $-\overline{uv}$. (a-c) Normalised by
$u^*=(-\overline{uv})^{1/2}$; (d) Normalised by $\overline{\tau}_{xy}$. The lines are as
indicated in table \ref{tb:parameters} (LB, LW, LWc and LN are presented by black, red, green and blue colors, respectively). In (a-c), solid lines with triangles are the
$Re_\tau=934$ channel by \cite{delAlamo04}. The linear-stress layer for LW and LWc is
indicated by gray shaded area and that for LN is indicated by vertical dotted lines. The
vertical scale is kept as in figure \ref{fig:stat_rms} to facilitate comparison.
} 
\label{fig:stat_rms_uv}
\end{figure}

The RMS velocity fluctuations in LWc are stronger than in LW. This is expected, because
previous investigations of channel flows with altered stress profiles
\citep{Tuerke13,Lozano-Duran19b} have concluded that the magnitude of the fluctuations
within the logarithmic layer scales with the local value of tangential Reynolds stress, and
because the primary role of turbulent fluctuations in the logarithmic layer is to carry
the tangential Reynolds stress required for the transfer of momentum. To check this, the RMS
velocity profiles are shown in figure \ref{fig:stat_rms_uv} scaled with the local velocity
scale $u^*=(-\overline{uv})^{1/2}$. Figure \ref{fig:stat_rms_uv}(d) confirms that it is
indeed true that most of the mean shear stress is carried by $-\overline{uv}$ within the
linear-stress layer, as in the logarithmic layer of natural flows. The profiles of LW and
LWc now agree well, but the consistent decrease with decreasing $\delta_a$ remains,
especially for $u'/u^*$. Note that figure \ref{fig:stat_rms_uv}  includes profiles from
the  DNS channel by \cite{delAlamo04}, whose $h^+=934$ is comparable to the $\delta_a^+$
of LW and LWc. The three flows agree reasonably well.

\begin{figure}
\centering
\includegraphics[width=0.5\linewidth]{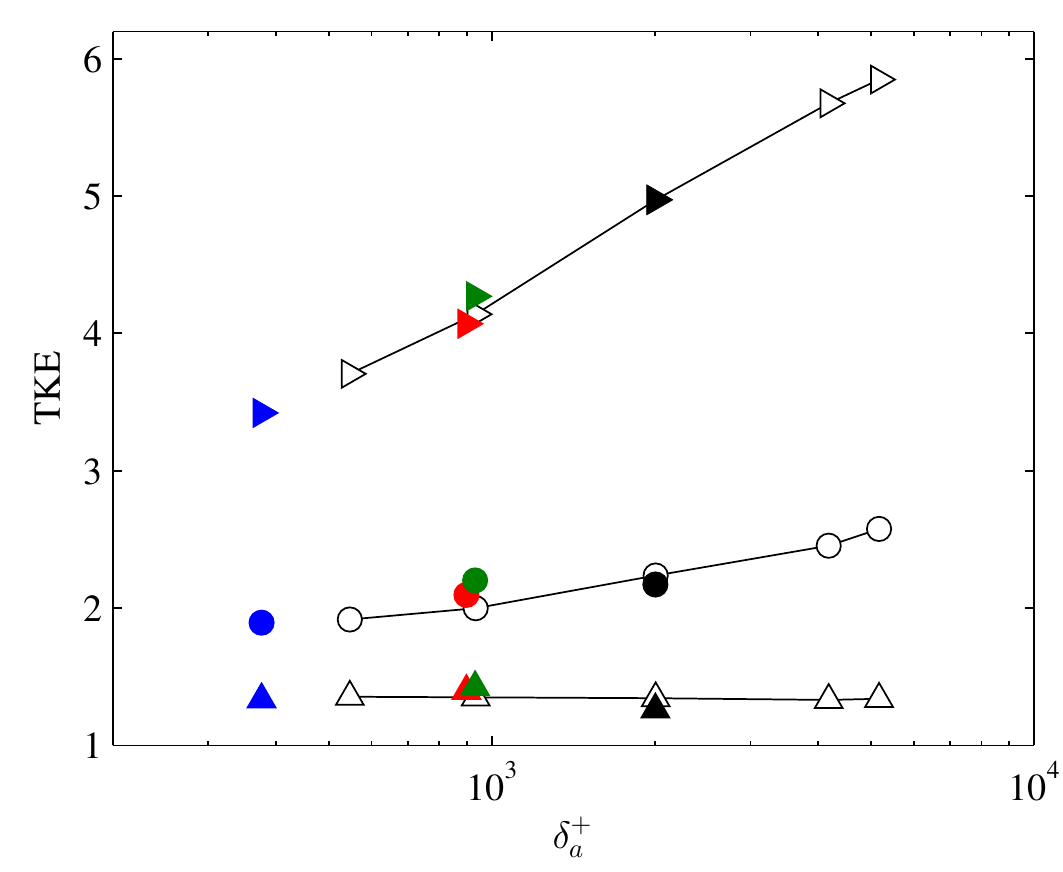}
\caption{TKE of each velocity component normalised by the local $\overline{uv}$ averaged
over $100<y^+<200$. $\rhd$: $\overline{u^2}/{-\overline{uv}}$, $\triangle$:
$\overline{v^2}/{-\overline{uv}}$, $\Circle$: $\overline{w^2}/{-\overline{uv}}$. Open symbols
connected with solid lines represent DNS databases at $Re_\tau=547$ \citep{delAlamo03}, 934
\citep{delAlamo04}, 2004 (HJ06), 4179 \citep{Lozano-Duran14a} and 5181 \citep{Lee15mk}.
Solid symbols represent LES experiments LB (black), LW (red), LWc (green) and LN (blue).}
\label{fig:rms_trend}
\end{figure}

We therefore turn our attention to the effect of $\delta_a^+$, and plot in figure
\ref{fig:rms_trend} the average value in $100<y^+<200$ of the TKEs of the three velocity
components, as functions of $\delta_a^+$ for different DNS databases and LES experiments.
The averaging range is chosen to be within the active or logarithmic layer in all the
datasets included, and we set $\delta_a=h $ for the DNS databases. In all cases, the TKEs
are normalised with the local $\overline{uv}$. For the DNS databases, $\overline{u^2}$ and
$\overline{w^2}$ display a log-linear trend with respect to $h^+$, while $\overline{v^2}$
stays roughly constant. This is consistent with the predictions from the attached-eddy
hypothesis \citep{Perry77,Perry82}, in which the main effect of increasing $h^+$ is
considered to be to extend the range of scales of the self-similar attached-eddy hierarchy.
The results from the LES experiments (solid symbols) agree well with the trend of the DNS
databases, except for a slight $\overline{w^2}$ excess for LW and LWc, which is due to the
mild hump in their $w'$ profile within the linear-stress layer (figure
\ref{fig:stat_rms_uv}c). Besides reinforcing the importance of $\delta_a$ as a parameter,
this agreement supports the equivalence of $\delta_a$ and $h$ in natural channels,
suggesting that the active part of the largest Townsend-type self-similar attached eddies
reaches the channel centreline, even though they are obscured in that region by the presence
of wake structures \citep[see also][]{delAlamo06,Lozano-Duran12}.
Another implication of this result is that the level of TKE in the logarithmic layer is almost exclusively determined
by the scale separation among the self-similar momentum-transferring eddies, while the
Reynolds shear stress  provides the velocity scale. In other words, for the isolated layers, $\delta_a$ acts as {a control parameter} that determines the scale separation as well as the mean shear as a function of $y/\delta_a$ (figure \ref{fig:stat_S}b), and $\delta_a$ is an independent parameter from the shear stress gradient, unlike natural channel flows. This is made especially clear by the
agreement between LW, LWc and \cite{delAlamo04} despite having different mean Reynolds shear
stress gradient and driving forces. Such comparisons are not possible in
natural channel flows because the scale separation within the self-similar attached eddies
and the mean stress gradient both depend on the Reynolds number.

\subsection{Spectra}

\begin{figure}
\centering
\includegraphics[width=0.8\linewidth]{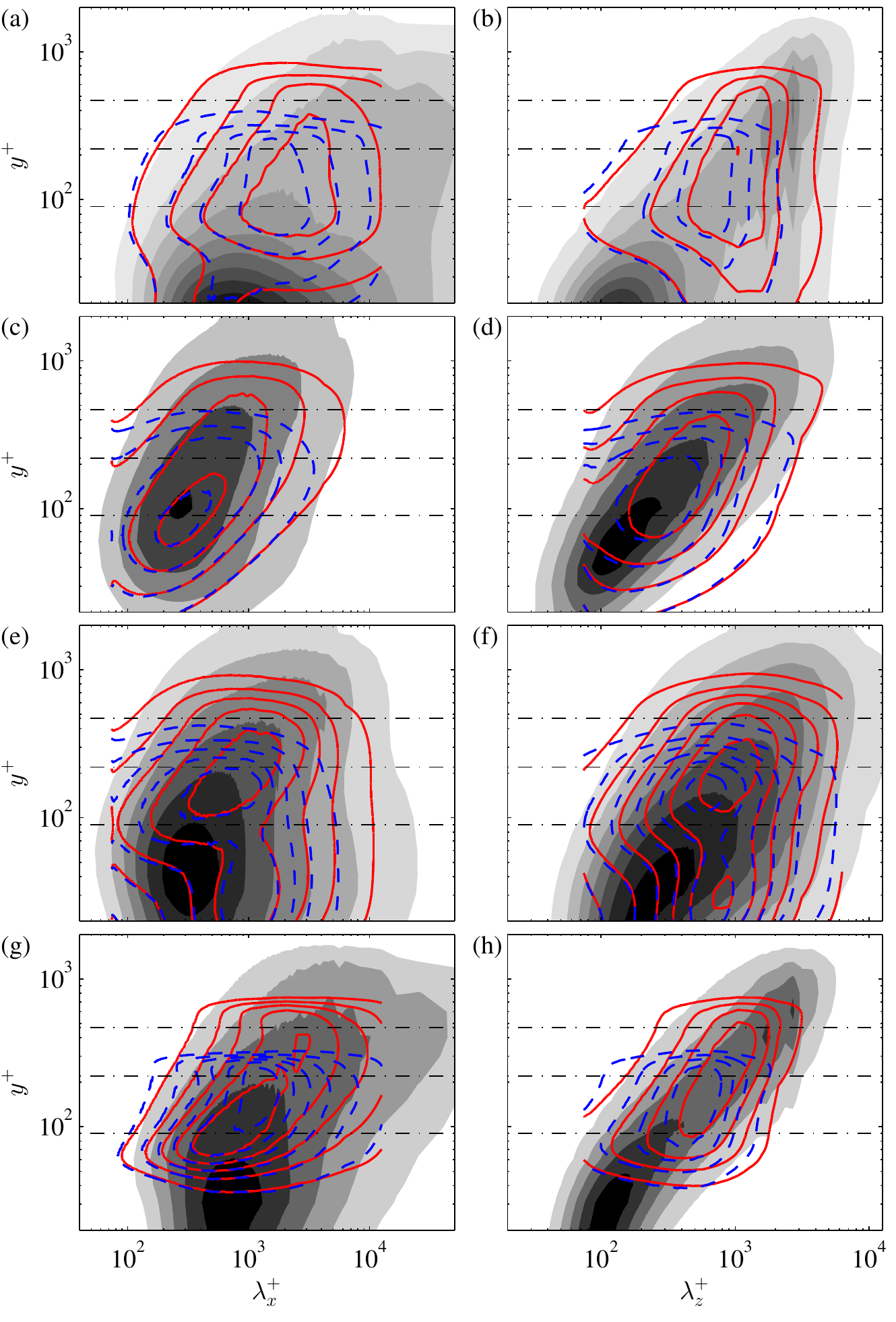}
\caption{One-dimensional pre-multiplied spectral density of (a,b) $u^2$; (c,d) $v^2$; (e,f)
$w^2$; (g,h) $-uv$ along the (a,c,e,g) streamwise; (b,d,f,h) spanwise direction. The gray
shaded contours are for HJ06, solid contours are for LW and dashed contours are for LN.
Contour lines are drawn at multiples of $0.1U_\tau^2$ except for (a) 0.2; (b) 0.4; (g) 0.05.
The horizontal dashed-dotted lines are $y/h=0.045$, $0.11$ and $0.235$, and mark the
boundaries of the linear-stress layer for LW and LN.}
\label{fig:spec_1D}
\end{figure}

To examine the distribution of turbulent kinetic energy at different scales, one-dimensional
premultiplied spectra are plotted in figure \ref{fig:spec_1D}. All the spectra are suppressed
outside the linear-stress layer, but the most notable observation is the elimination of the
near-wall spectral peak in the spectrum of $u$ for LW and LN, which is especially clear in
figure \ref{fig:spec_1D}(b) and proves that our numerical experiment effectively
removes the dynamics of the buffer layer. Another important difference is the
attenuation, within the linear-stress layer of LW and LN, of the spectrum of $u$ at very large
$\lambda_x$ and $\lambda_z$. The motions in this range of wavelengths are commonly referred
to as the very large-scale motions (VLSM) of the logarithmic layer \citep[e.g.][]{Jimenez98,Kim99}
and some attention has been dedicated to them, because they carry a substantial
fraction of the TKE and of the Reynolds stresses \citep{Balakumar07}. However, the present
result suggests that the VLSMs are not part of the intrinsic dynamics of the
logarithmic layer, but of the region above it, which has been suppressed by the body force in LN and LW. This
idea is consistent with the concepts of `inertial waves'  in \cite{Jimenez18} or of `global
modes' in \cite{delAlamo03}, introduced to describe the energetic motions of $u$ at very large
wavelengths, and which occupy the majority of the channel half width. \cite{Kwon_thesis} tried a different way of eliminating the outer layer contributions to the
velocity fluctuations, from the perspective of the quiescent core. He observed that, upon the
removal of the velocity fluctuations associated with the quiescent core, most of the energy
of $u$ in the VLSM range disappears.

\begin{figure}
\centering
\includegraphics[width=0.8\linewidth]{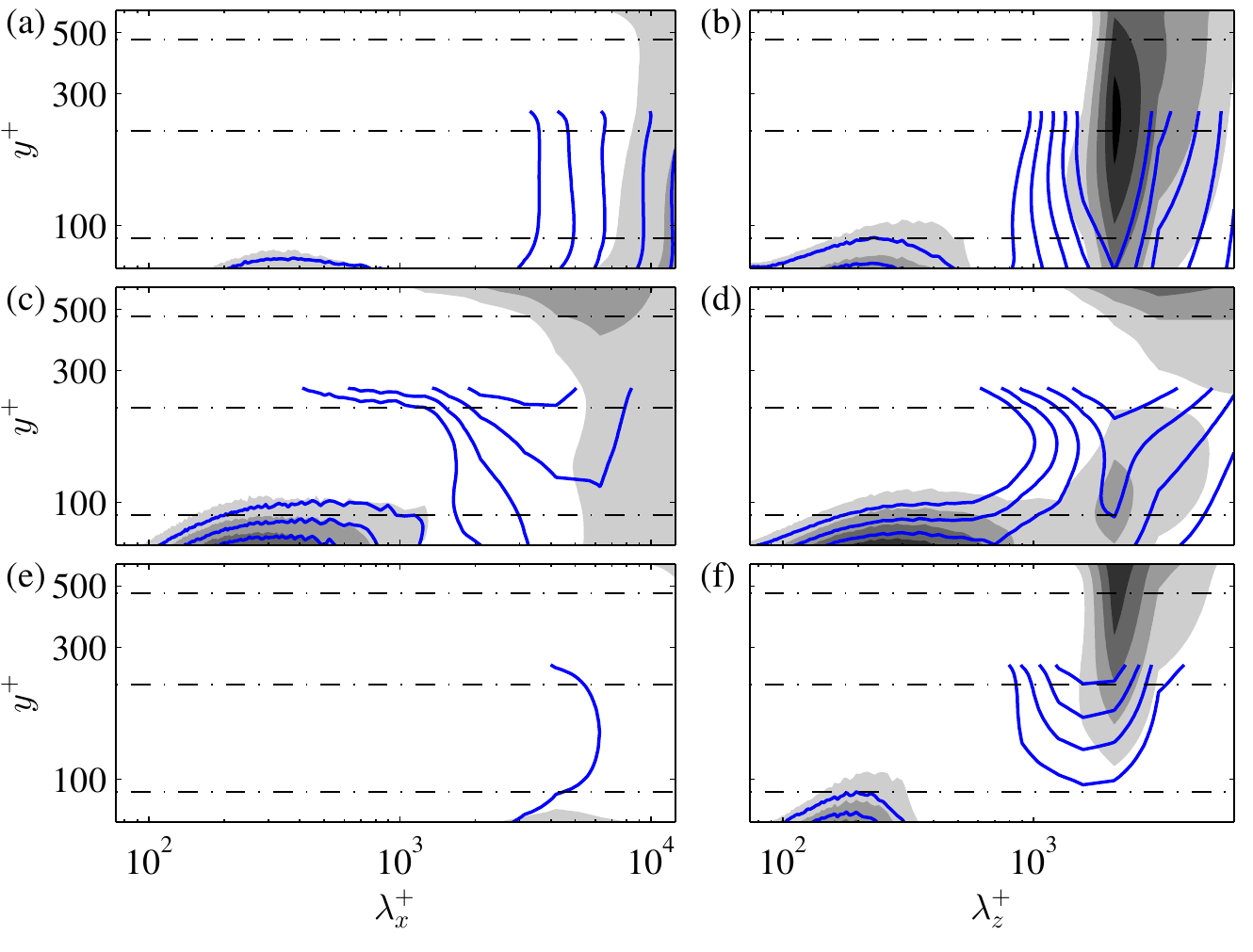}
\caption{Difference between the one-dimensional pre-multiplied spectral density of (a,b)
$u^2$; (c,d) $w^2$; (e,f) $-uv$ along the (a,c,e) streamwise; (b,d,f) spanwise direction.
The grey shaded contours are LB-LW and lines contours are for LB-LN. In (a) the contours are
separated by $0.1 U_\tau^2$. In (b), by $0.2 U_\tau^2$. In (c-f), by $0.05 U_\tau^2$. The
horizontal lines are the limits for the two linear-stress layers.
}%
\label{fig:spec_1Ddiff}
\end{figure}

The damping of the long and wide wavelengths in LN and LW is made explicit in figure
\ref{fig:spec_1Ddiff}, which shows the difference between their spectra and the full LES
case. The restricted layers exhibit an energy deficit with respect to LB, and it is restricted to the large
scales. Moreover, the length of the region in which LN falls below LB (e.g.
$\lambda_x^+\approx 3500$ at the $0.1U_\tau^2$ level of $k_x\phi_{uu}$ in figure
\ref{fig:spec_1Ddiff}a) is approximately twice shorter than for LW, proportionally to their
respective $\delta_a$. The width of the spanwise defect follows a similar trend but the peak
is located at $\lambda_z \approx h$ in both cases, consistent with the known width of the
VLSM \citep{Jimenez18}. Note that there are no plots for $\phi_{vv}$ in figure
\ref{fig:spec_1Ddiff}. This velocity component has no VLSM, and the corresponding plots are
almost empty.
 
All these studies converge to the conclusion that the VLSMs do not belong to the
self-similar wall-attached eddy hierarchy intrinsic to the logarithmic layer. This is not to
say that they have no influence on its dynamics, but it suggests
that the origin and dynamics of the VLSMs are associated with the outer layer rather than with
the logarithmic one. A similar attenuation of the large scales is observed for $w$ and, to a
lesser degree, for $uv$, but not for $v$, consistent with \citeauthor{Townsend76}'s
(\citeyear{Townsend76}) idea that the $u$ and $w$ fluctuations are attached, in the sense that
they are created far from the wall and extend downwards to fill the space underneath, while
the $v$ fluctuations are local in $y$. This is also clear from the triangular spectral
`skirts' in figures \ref{fig:spec_1D}(a,b) and \ref{fig:spec_1D}(e,f). That these roots are
`inactive' with respect to the tangential stress is shown by the lack of skirts in figures \ref{fig:spec_1D}(c,d) and
\ref{fig:spec_1D}(g,h). It is interesting to observe the dependence of the large-scale
energy attenuation on the thickness of the linear-stress layer, demonstrated by the greater
attenuation in LN compared to LW. This supports the idea that restricting the wall-normal
dimension over which turbulent fluctuations can develop also limits their growth in the wall-parallel
directions. Long structures at a given $y$ are the skirts of structures farther up, and
truncating the top of the spectral triangle, also truncates the long wavelengths. In other words, the
structures in the linear-stress layer are `minimal' in the wall-normal direction, and the
upper bound of linear-stress layer acts as a `ceiling' that limits the growth of the structures
in the wall-parallel directions as well. This also explains the decreasing trend of $u'$ and
$w'$ with decreasing $\delta_a$ in figure \ref{fig:rms_trend}.

\begin{figure}
\centering
\includegraphics[width=0.8\linewidth]{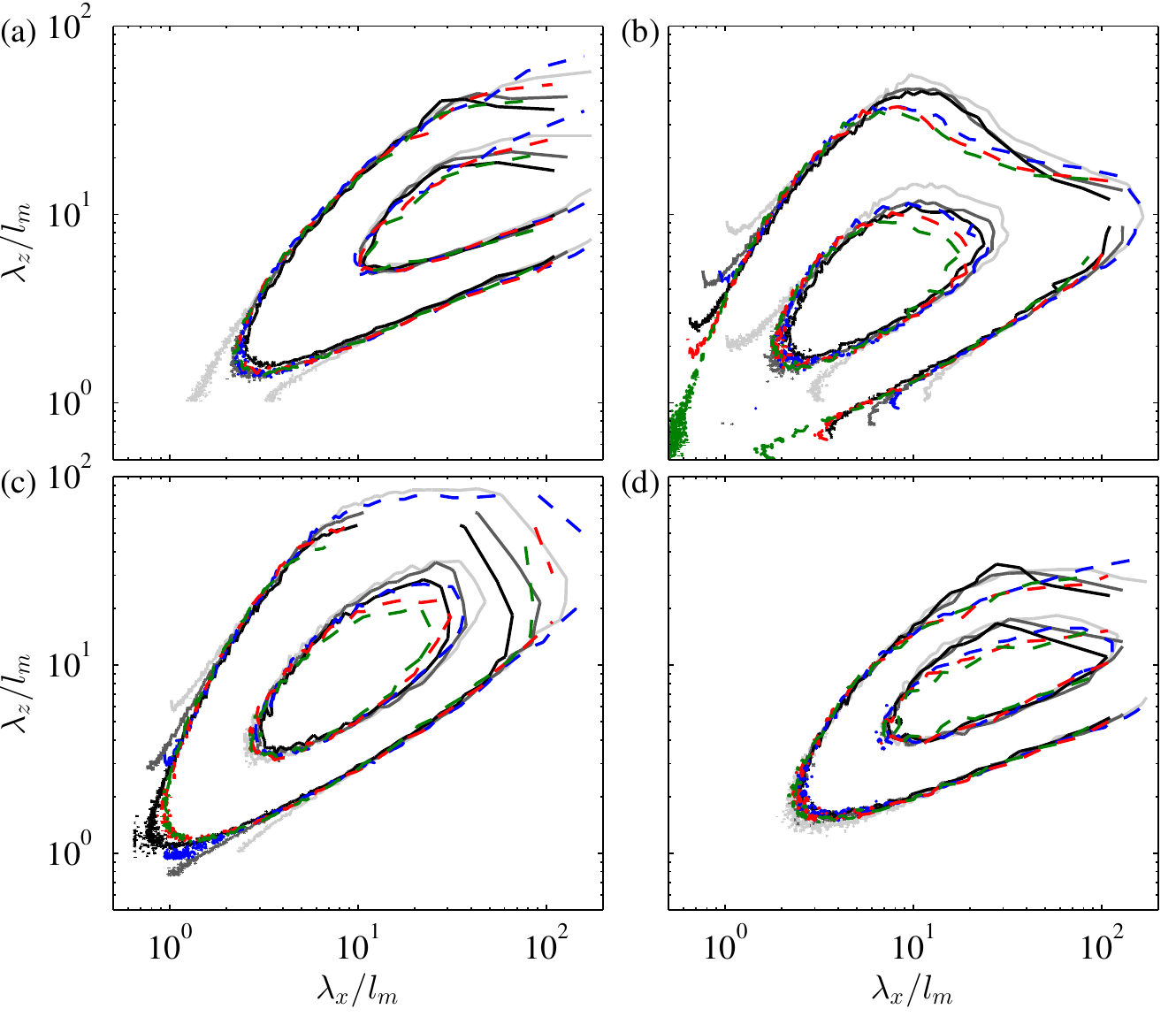}
\caption{Contour plots of (a) $k_xk_z\phi_{uu}$; (b) $k_xk_z\phi_{vv}$; (c)
$k_xk_z\phi_{ww}$; (d) -$k_xk_z\phi_{uv}$ against $\lambda_x/l_m$ and $\lambda_z/l_m$,
scaled by the mixing length at each height. The solid grey scale contours are for LW and
computed at $y/h\simeq0.1$, 0.15 and 0.2 (from light gray to black). The dashed color
contours are for LB and computed at $y/h\simeq0.1$ (blue), $0.15$ (red) and $0.2$ (green).
Contour levels are drawn at (a) $[0.1~ 0.3]U_\tau^2$; (b,d) $[0.03~0.1]U_\tau^2$; (c)
$[0.05~0.15]U_\tau^2$.}
\label{fig:spec_2D}
\end{figure}

Figure \ref{fig:spec_2D} presents two-dimensional velocity spectra at three wall-normal
locations to examine the self-similarity of the velocity fluctuations in the isolated
linear-stress layer, which is the defining characteristic of natural logarithmic layers. For
brevity, only LW and LB are compared in the figure, but LN and LWc display a similar
collapse. The wavelengths are scaled with the local mixing length, which was shown by
\cite{Mizuno11} to collapse the velocity spectra in DNS channels better than the distance
from the wall. More recently, \cite{Lozano-Duran19b} proposed a similar length scale, also
based on the mean shear but using the local velocity scale $\overline{-uv}^{1/2}$ instead of
$U_\tau$, but the difference between the two scales is small for the range of wall-normal
locations in figure \ref{fig:spec_2D}, and we have kept the traditional definition. The
energetic cores of the spectra of LW at different wall-normal locations show an excellent
collapse, supporting the conclusion that the mixing length is the correct length scale for
the energy-containing eddies in the logarithmic layer, even when the profile of the mixing length is not linear.
{If the typical velocity scale within the logarithmic layer is given by $U_\tau$}, this implies that the time scale of the {energy-containing} eddies is dictated by the local mean shear, rather than by a local eddy turnover based on the distance from the wall {and $U_\tau$}.
The core of the spectra of LW also agree well with LB. The lack of collapse at the
large-scale ends of $k_xk_z\phi_{uu}$ and $k_xk_z\phi_{ww}$ was already discussed in figure
\ref{fig:spec_1D}, and corresponds to the inactive structures, which scale with $h$ or with
$\delta_a$. In particular, note the damping of the spectrum of LW in the upper-right
corner of figure \ref{fig:spec_2D}(a,d).

\begin{figure}
\centering
\includegraphics[width=0.8\linewidth]{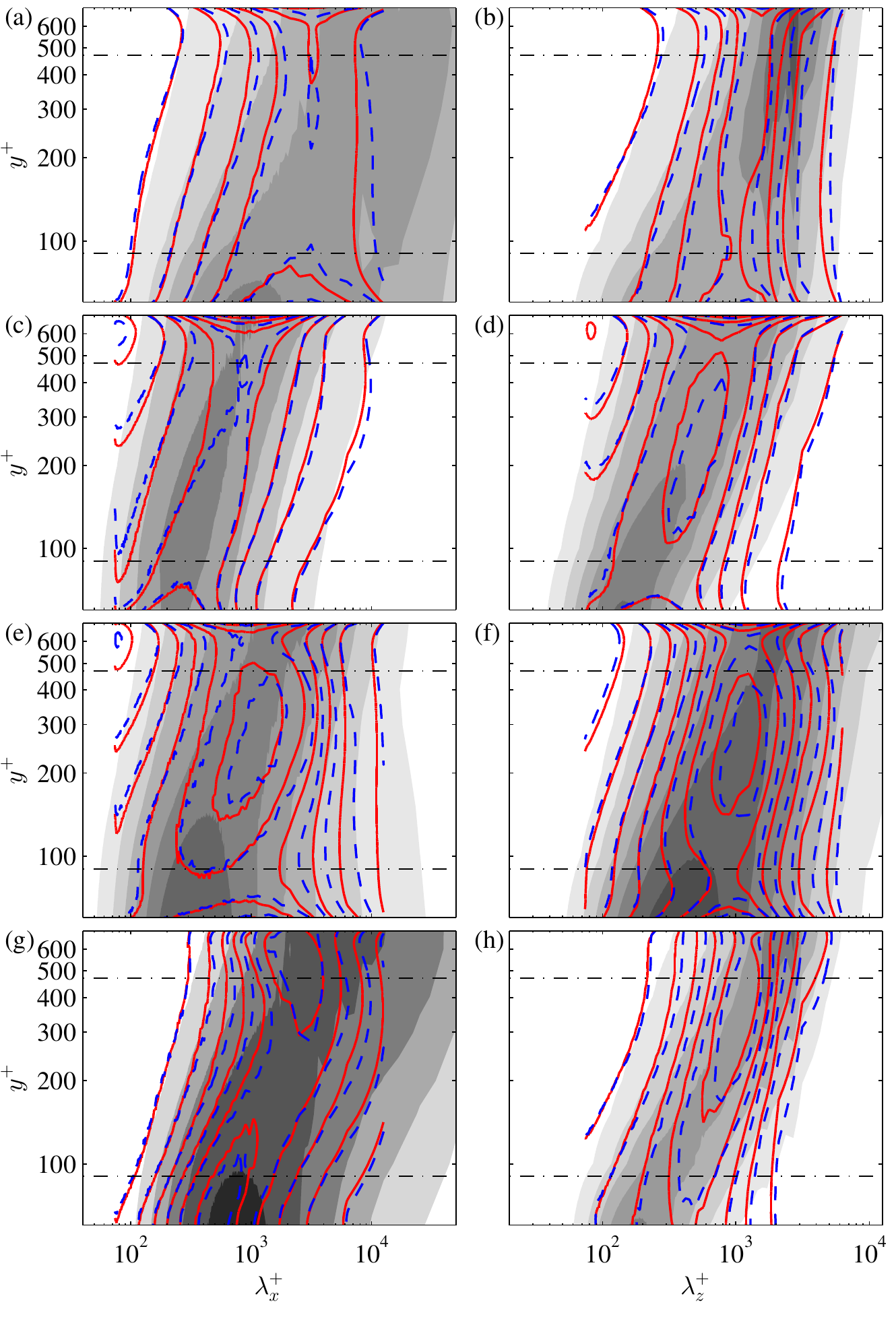}
\caption{One-dimensional pre-multiplied spectral density of (a,b) $u^2$; (c,d) $v^2$; (e,f) $w^2$; (g,h) $-uv$ along the (a,c,e,g) streamwise; (b,d,f,h) spanwise direction. The gray shaded contours are for HJ06, solid contours are for LW and dashed contours are for LWc. Contour lines are drawn at multiples of $0.1u^{*2}$ except for (a) 0.2; (b) 0.4; (g) 0.05. The horizontal dashed-dotted lines indicate $y=0.045h$, and $0.235h$, which mark the boundaries of linear-stress layers.}
\label{fig:spec_1D_uv_scaling}
\end{figure}

\begin{figure}
\centering
\includegraphics[width=0.8\linewidth]{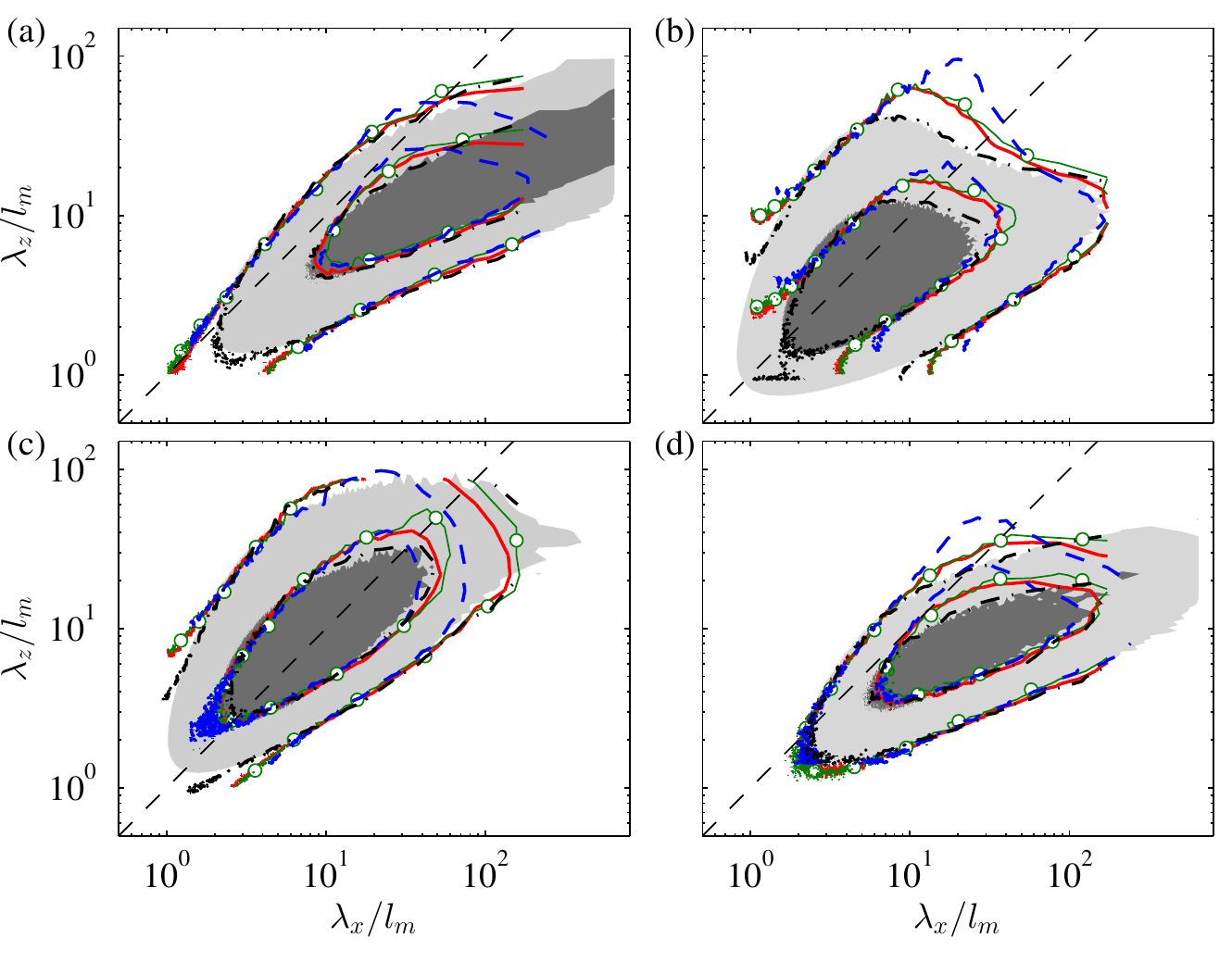}
\caption{Contour plots of (a) $k_xk_z\phi_{uu}$; (b) $k_xk_z\phi_{vv}$; (c) $k_xk_z\phi_{ww}$; (d) -$k_xk_z\phi_{uv}$ at $y\simeq0.1h$ plotted against $\lambda_x/l_m$ and $\lambda_z/l_m$. The shaded contours are HJ06. The line contours are LES experiments as indicated in table \ref{tb:parameters} (LB, LW, LWc and LN are presented by black, red, green and blue colors, respectively). Contour levels are drawn at (a) $[0.1~ 0.3]u^{*2}$; (b,d) $[0.03~0.1]u^{*2}$; (c) $[0.05~0.15]u^{*2}$. The black dashed diagonal lines represent $\lambda_x=\lambda_z$.}
\label{fig:spec_2D_comp}
\end{figure}

Figure \ref{fig:spec_1D_uv_scaling} examines the effect of changing the stress profile in
layers of similar thickness by comparing cases LW and LWc. The one-dimensional spectra are
normalised with $u^*$, which was shown in the previous section to be the correct scale for
the intensities, and shown only within the linear-stress layer. They collapse well, showing
that the spectral distribution of the fluctuations, and not only their TKE, is independent
of the existence of a pressure gradient.

To complement the observations on the trend of the large-scale energy attenuation, figure
\ref{fig:spec_2D_comp} compares two-dimensional energy spectra at $y/h\simeq0.1$, scaled
with $u^*$. The spectra for the full LES (LB) agree well with HJ06, again demonstrating the
adequacy of the current LES simulations for the study of the logarithmic layer. There is
some accumulation of energy at the scales close to the grid resolution of LES, due to the
slightly insufficient dissipation by the SGS model, but this effect does not extend to the
energy-containing region. The spectra for LW and LWc agree well, reinforcing the conclusions
from the one-dimensional data. The comparison between LB, LW and LN clearly shows the
removal of large-scale energy as the width of the linear-stress layer decreases, especially
for $u$ and $w$. This also explains the previously observed decreasing trend of the $u'$ and
$w'$ profiles with decreasing $\delta_a^+$, discussed in figure \ref{fig:stat_rms}.

On the other hand, there is a TKE excess in the spectra of all the restricted-layer
experiments with respect to LB and HJ06, at intermediate scales, which can be observed in
figures \ref{fig:spec_2D} and \ref{fig:spec_2D_comp}. The wavelengths of the excess energy
in LW scale with the mixing length above $y/h=0.075\,(y^+=150)$, and are centred around
$(\lambda_x,\lambda_z) \simeq (15l_m,10l_m)$ for $v$ and $w$, and around
$(\lambda_x,\lambda_z) \simeq (30l_m,10l_m)$ for $u$. This energy excess also produces some
extra Reynolds shear stress, which appears as an upper `horn' in figure
\ref{fig:spec_2D_comp}(d), and which is needed to compensate for the attenuated Reynolds
shear stress at the large scales. The same energy excess also appears in LN, where it is
stronger because it has to compensate for an even larger attenuation. Except for these
localised effects, the agreement in other regions of the two-dimensional spectra is very
good.

There is also an  energy excess in the wide modes of $v$ in LN, which can be observed in
figure \ref{fig:spec_2D_comp}(b), and which also appears in LW near the lower limit of the
linear-stress layer (not shown). Because its wavelengths are wide and relatively short, it
is tempting to attribute this extra energy to a spanwise instability of the shear layer that
forms underneath the linear-stress region (see the peak at $y^+\approx 70$ in figure
\ref{fig:stat_S}). The extra energy in the near-wall region is also visible in figure
\ref{fig:spec_1D}(a,b) as a `stem' at $\lambda_x^+\approx 700$ and $\lambda_z^+\approx
2000-3000$, hanging below the spectrum of LW into the near-wall damped layer, and it can
also be shown that the spectrum of $u$ at $y^+=20$, although weak overall, contains very
wide structures with $\lambda_x^+\approx 700$. Note that these wavelengths are much wider in
the spanwise direction than any residual near-wall peak that might have not been fully
damped by the forcing. \cite{jim:uhl:pin:kaw:01} showed that any profile with an approximate
inflexion point near the wall develops a Kelvin--Helmholtz like instability as soon as any
amount of wall transpiration is allowed, and \cite{gmayjim11} showed that, in ribbed
surfaces modelled by a layer of retarding body forces, this effect results in transverse unstable
rolls. This instability, like Kelvin--Helmholtz's, is essentially inviscid and depends only
on the mean $u$ profile,  and in the $v$ velocity that separates the inflection point from the
impermeable wall. Its typical wavelengths are of the order of 5 to 10 times the thickness of
the drag layer, which in the present case $(\lambda_x^+\approx 450-900)$ would not be too
far from the observed values. In the case of the ribbed surface mentioned above, the effect
of the instability is mostly confined to the damped layer. However, in our cases, there is
another shear layer above the linear-stress layer, and there could be a resonance between two
shear layer instabilities located below and above the linear-stress layer, which may act as
a seeding mechanism for the energy excess within the isolated layer. To confirm this
possibility is beyond the scope of the present paper, but the direct influence of the
possible instabilities does not seem to be significant, and the characteristics of the
energy-containing eddies are well-reproduced.

\subsection{Dynamical indicators of the flow}

Examination of velocity statistics reveals that our numerical experiment can replicate key
kinematic properties of the natural logarithmic layer. In order to further assess the
resemblance of the linear-stress layer to the natural logarithmic layer, we also examine and
compare some of the dynamical characteristics of the flow. Firstly, we will examine the
ratio between the production and dissipation of TKE, since it is widely known that these two
quantities are approximately in balance in the logarithmic layer. For LES channel flows, the
balance of the mean TKE of the filtered velocity fields is given by
\begin{equation}
\frac{Dk}{Dt} = P+\epsilon+\Pi_p+\Pi_t+\Pi_v+\Pi_r \label{eq:budget}
\end{equation}
where the terms in the right-hand side represent production, dissipation, pressure transport, turbulent transport, viscous diffusion and diffusion by residual stress, respectively. They are defined as
\begin{align}
P &=-S\overline{uv} \\
\epsilon &= -\overline{\mathbf{\tau^v}:\mathbf{E}}+\overline{\mathbf{\tau^r}:\mathbf{E}} \\
\Pi_p &= -\frac{\partial\overline{vp}}{\partial y} \\
\Pi_t &= -\frac{\partial\overline{vk}}{\partial y} \\
\Pi_v &= \nabla \cdot \overline{\mathbf{u}\cdot\mathbf{\tau^v}}\\
\Pi_r &= -\nabla \cdot \overline{\mathbf{u}\cdot\mathbf{\tau^r}}\\
\end{align}

\noindent where $\mathbf{u}$ is the fluctuating velocity vector $(u,v,w)$, $p$ is the kinematic
pressure, $\mathbf{E}$ is the strain rate tensor of the filtered velocity fields,
$\mathbf{\tau^v}=2\nu\mathbf{E}$ is the viscous stress tensor and
$\mathbf{\tau^r}=-2\nu_r\mathbf{E}$, where $\nu_r$ is the LES eddy
viscosity given by the SGS model, is the residual stress tensor.

\begin{figure}
\centering
\includegraphics[width=0.8\linewidth]{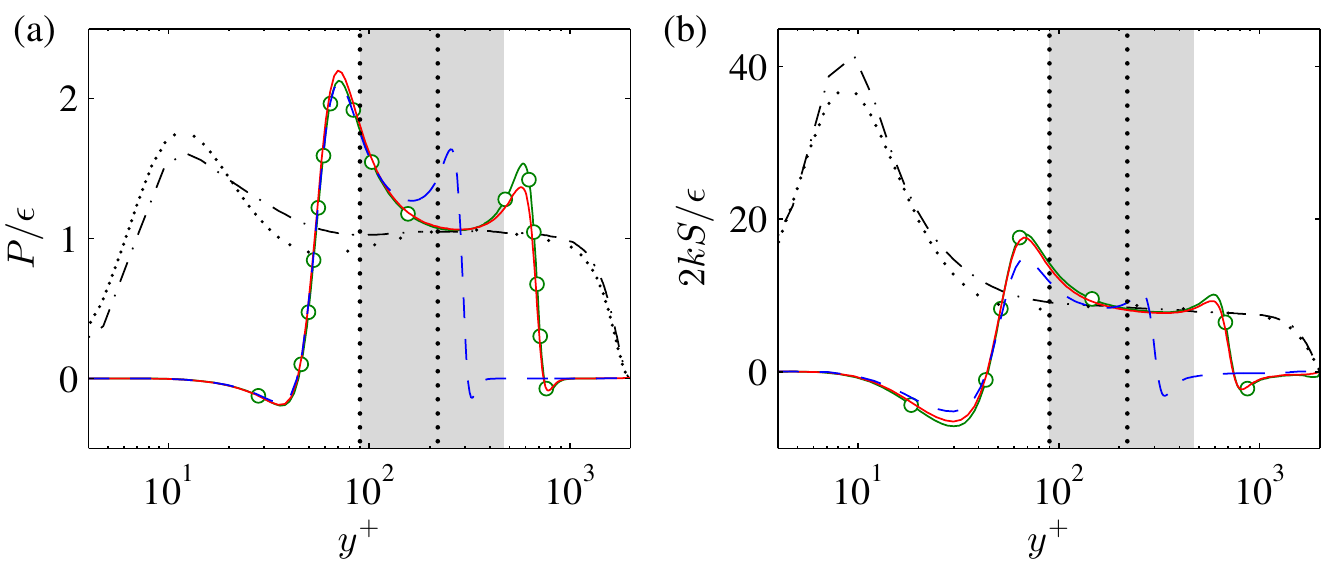}
\caption{(a) Ratio between production and dissipation (b) Corrsin parameter. The lines are as indicated in table \ref{tb:parameters} (LB, LW, LWc and LN are presented by black, red, green and blue colors, respectively). The linear-stress layer for LW and LWc is indicated by gray shaded area and that for LN is indicated by vertical dotted lines.} 
\label{fig:balance_corrsin}
\end{figure}

Another key parameter for characterizing the dynamics of shear flow is the Corrsin
parameter, which is defined as the ratio between the dissipative time scale of eddies
($2k/\epsilon$) and the time scale of the mean shear ($1/S$). Therefore, it represents the
relative importance of shear to the dynamics of turbulent eddies. For example,
$2kS/\epsilon\gg1$ means that the structures live long enough to experience the effects of
the shear. Figure \ref{fig:balance_corrsin} shows the plots of the production/dissipation
ratio and the Corrsin parameter for all the LES cases and for HJ06. The two full channels, LB
and HJ06, show some discrepancies below $y^+\simeq100$, but agree reasonably well above that
level, demonstrating the adequacy of our LES scheme for studying the logarithmic layer. For
these two flows, $P/\epsilon \simeq 1$ over a wide range of wall-normal locations
corresponding to the conventional logarithmic layer. The two isolated layers LW and LWc,
have a narrower region in which the deviation of the $P/\epsilon$ profile with respect to LB
remains less than $5\%$, located in $200\lesssim y^+\lesssim390$ ($0.1\lesssim
y/h\lesssim0.195$) . On the other hand, $P/\epsilon$ ratio never reaches unity in LN,
presumably because the linear-stress layer is too narrow to fully recover the TKE balance.
In the middle of the linear-stress layer, the Corrsin parameter in figure
\ref{fig:balance_corrsin}(b) agrees well among all the LES cases and HJ06, and stays
approximately constant at $2kS/\epsilon\approx 8.$ This agreement suggests that similar
dynamical processes take place in all these flows. These observations demonstrate that our
numerical experiments are able to produce a region whose dynamical characteristics are similar
to those of the natural logarithmic layer.

\begin{figure}
\centering
\includegraphics[width=0.8\linewidth]{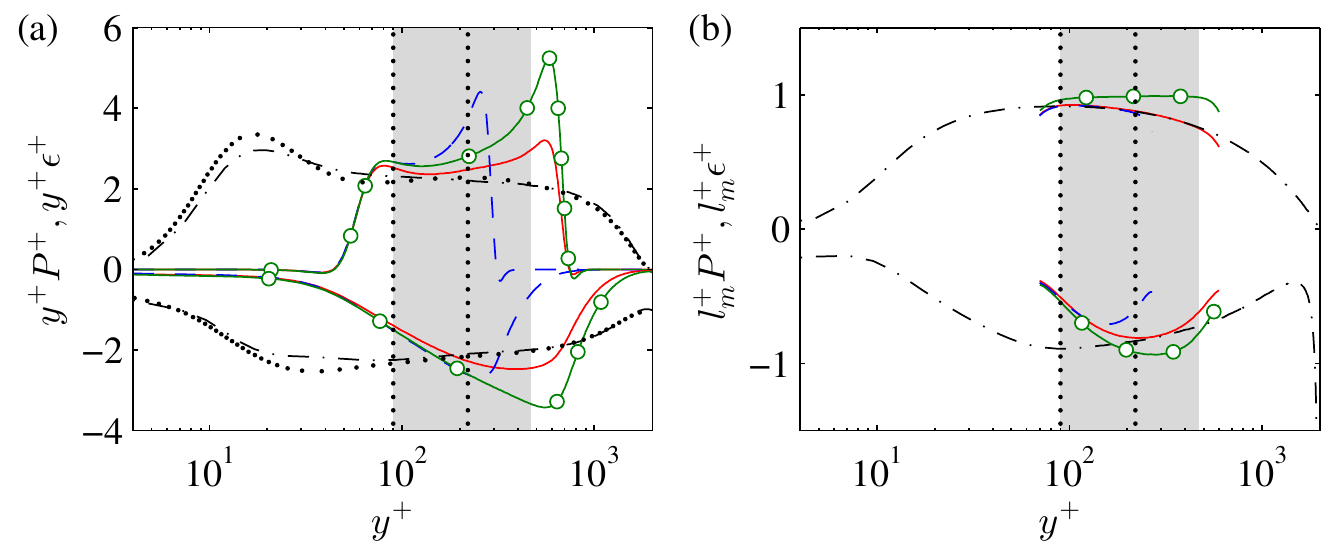}
\caption{Profiles of production and dissipation. Lines on the positive and negatives sides represent the production and dissipation, respectively. (a) Premultiplied by $y$ (b) Premultiplied by $l_m$. The lines are as indicated in table \ref{tb:parameters} (LB, LW, LWc and LN are presented by black, red, green and blue colors, respectively). The linear-stress layer for LW and LWc is indicated by gray shaded area and that for LN is indicated by vertical dotted lines. In (b), profiles for LW and LWc are plotted only near the vicinity of the linear-stress layer ($70<y^+<600$).}
\label{fig:prod_dissip}
\end{figure}
 
However, the profiles of $P/\epsilon$ and of the Corrsin parameter for LW have peaks at
$y^+\simeq70$ and $y^+\simeq570$ in figure \ref{fig:balance_corrsin}, just outside of the
linear-stress layer. This could potentially be worrying, because the excess $P/\epsilon$ in
these regions may influence the dynamics of the linear-stress layer. To investigate this
possibility, figure \ref{fig:prod_dissip} compares the actual production and dissipation
profiles. When $P$ and $\epsilon$ are premultiplied by $y$ in figure
\ref{fig:prod_dissip}(a), to highlight the logarithmic and outer regions, they do not agree
well, even within the linear-stress layer. However, we saw in the previous section that the
correct length scale for this region is the mixing length, and when the production is
premultiplied by $l_m$ in figure \ref{fig:prod_dissip}(b), LW, LN and LB agree excellently,
while LWc does not. This is essentially automatic, because $l_m^+P^+=-\overline{uv}^+$,
which is set by the body force, whose profile is only different for LWc. The behaviour of
$l_m^+\epsilon^+$ is more interesting. The agreement between LB and LW near $y^+\simeq280$
is consistent with the collapse of $l_m^+P^+$ and of $P/\epsilon$ in this region, but it is
clear from figure \ref{fig:prod_dissip}(b) that the peaks of $P/\epsilon$ near the edge
of the linear layer are caused by a reduced level of dissipation, not by an increased level
of production. The balance of the two quantities is never reached for LN, because its
linear-stress layer is too narrow (the scale separation between the two edges is only a
factor of 2). \cite{Tuerke13} investigated a turbulent channel flow with a sharp change in
the mean shear, and found that the production adapts to the change in the shear almost
immediately, while the dissipation does so more gradually, in agreement with the behaviour
near the edges of the linear-stress layer in figure \ref{fig:prod_dissip}(b). They
attributed this phenomenon to the temporal delay between the production and dissipation
mechanisms.

To further inspect this behaviour, the wall-normal flux of the mean TKE is considered.
The transport terms in (\ref{eq:budget}) are in the form of a flux
divergence, and the wall-normal flux of the mean TKE, $\Theta$, can be computed as
\begin{equation}
\Theta(y)=\int_{0}^{y}\left(\Pi_p(\xi)+\Pi_t(\xi)+\Pi_v(\xi)+\Pi_r(\xi)\right)\dd \xi .
\label{eq:theta}
\end{equation}

\begin{figure}
\centering
\includegraphics[width=0.6\linewidth]{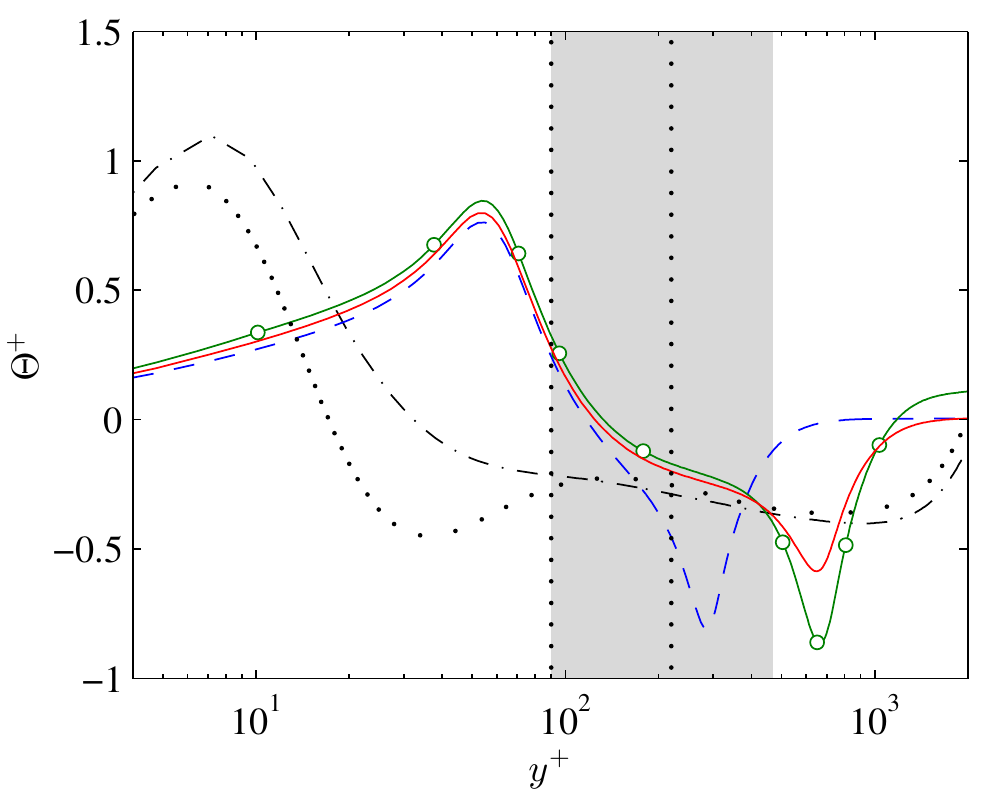}
\caption{Profiles of the wall-normal TKE flux. The lines are as indicated in table \ref{tb:parameters} (LB, LW, LWc and LN are presented by black, red, green and blue colors, respectively). The linear-stress layer for LW and LWc is indicated by gray shaded area and that for LN is indicated by vertical dotted lines.}
\label{fig:TKE_flux}
\end{figure}

\noindent As per (\ref{eq:budget}), regions with positive $\dd \Theta/\dd y$ (i.e. positive
transport) indicate net energy sinks, $\epsilon>P$, which draw energy from other wall-normal
locations, and vice versa. Also, because (\ref{eq:theta}) vanishes at the wall, a positive
$\Theta$ indicates a net TKE flux towards the wall at that wall-normal location. Figure \ref{fig:TKE_flux} shows $\Theta$ 
for {the LES cases} and HJ06 \citep{Hoyas08}. {LW, LWc and LN agree well below the linear-stress layer but LN does not exhibit a plateau region because it never recovers the local TKE balance. LW, LWc, LB and HJ06 exhibit a mild plateau region within the linear-stress or logarithmic layer but} they only approximately agree in the upper half of the
linear-stress layer. The slope
of $\Theta$ in the logarithmic and linear-stress layers is negative, since $P/\epsilon$ is
slightly greater than unity there \citep[at least up to $Re_\tau=5200$;
see][]{Bernardini14,Lee15mk}. Otherwise, $\Theta$ has a positive slope at $y^+<55$ and $y^+>645$ for
LW, indicating that the flow  acts as an energy sink outside of the linear-stress layer, except in
the vicinity of the layer edges. In full channels, like LB and HJ06, the
buffer layer ($5<y^+<40$) acts as a strong energy source due to intense production
activities. However, the removal of the buffer layer in LW turns this region into a net
energy sink, whose energy deficit is balanced by an energy flux coming from the
linear-stress layer. The region above $y^+\simeq 645$ also acts as a net energy sink and
draws energy from the linear-stress layer. In LW and LWc, $\Theta$ crosses zero at $y^+\simeq125$,
meaning that the TKE gets transported towards the wall at $y^+<125$ and towards the centre
at $y^+>125$. On average, some of the TKE produced in the linear-stress layer gets
transported away from the layer, instead of being dissipated within it. This is different from
the behaviour of natural logarithmic layers, and explains the reduced level of dissipation
near the edge of the linear-stress layer. For $200<y^+<400$, the slope of $\Theta$ in LW
approaches that of LB and HJ06, while the actual magnitude of TKE flux is smaller because
there is less upward TKE flux coming from below. This region coincides with the
location where a good agreement is observed for $P/\epsilon$ and the Corrsin parameter
between LW, LWc and LB. Overall, the shear production mechanism of the logarithmic layer is well
reproduced in the linear-stress layer, although there are some differences in how the TKE is
transported and dissipated near the layer edges. Therefore, the current numerical experiment
is an adequate reproduction of the natural logarithmic layer as far as the energy producing
and energy containing motions are concerned.

\section{Discussion}

To isolate the logarithmic layer of wall-bounded turbulent flows, we have presented a series
of numerical experiments in which a body force is used to impose a prescribed total stress
profile. The resulting flow has an `active' layer in which the total stress follows a linear
trend, as in natural channels, but the stress is made to decay to zero elsewhere. As a
result, the turbulent fluctuations are effectively eliminated outside the active layer,
especially the ones carrying the tangential Reynolds stress. Various statistical comparisons
demonstrate the kinematic and dynamic equivalence between the isolated active layer and the
natural logarithmic one, and the experiments allow us to assess separately the effects of
the range of scales of the self-similar eddies (using cases LB, LW and LN), and of the
profile of the shear stress within the active layer (using LW and LWc). These effects cannot
be separated in natural channel flows, because both are controlled by the Reynolds number.
We show that the scale range of the self-similar eddies is related to the thickness,
$\delta_a$, of the active layer, which controls the size of the largest
momentum-transferring eddies. This thickness determines the mean shear below
$y=0.4\delta_a$, and the largest wall-parallel scales of the flow. On the other hand, the
primary effect of the average shear stress within the linear-stress layer is to act as a
scale for the velocity fluctuations, while the slope of the shear stress profile, which is
equivalent in natural channels to the mean streamwise pressure gradient, does not
significantly affect the dynamics.

The characteristics of the energy-containing eddies are investigated using their energy
spectrum. Within the linear-stress layer, our experiments agree with natural channel flows
when the wavelengths are scaled with the mixing length of the mean streamwise velocity
profile. This agreement includes the spectrum at different wall-normal locations, even if
the mixing length profile is not strictly linear in $y$. This suggests that the linear dependence
of the length scale in natural channels is not a necessary condition for self-similarity,
and that the length scale of the self-similar eddies in the logarithmic layer is associated
with the local value of mean shear, not with the {absolute} distance from the wall, in line with the
conclusions of \cite{Mizuno11} and \cite{Lozano-Duran19b}. {The implication is that the distance from the wall relative to the size of the largest active eddies determines the mean shear (figure \ref{fig:stat_S}), and the mean shear, together with the mean momentum flux (roughly $U_\tau^2$ in the logarithmic layer), determines the scale of the self-similar eddies in the logarithmic layer, rather than the absolute distance from the wall. In this regard, although the absolute distance from the wall does not provide a scale for the self-similar active eddies, the isolated layer is not truly independent of the wall because the value of the mean shear depends on $y/\delta_a$.} Another noteworthy difference between
our numerical experiments and the natural channel is the attenuation in the former of the
turbulent kinetic energy (TKE) of the very large-scale motions, suggesting that these
motions are not an  intrinsic  part of the dynamics of the logarithmic layer.

By comparing experiments having linear-stress layers of different thickness, we confirm that
the upper bound of the layer acts like a `ceiling' for the structures, and that inhibiting
the wall-normal growth of turbulent structures also limits their wall-parallel scales. This
attenuation of the large-scale energy also explains the observed decrease of $u'$ and $w'$
as $\delta_a$ decreases, or equivalently, as the scale separation within the self-similar
eddy hierarchy gets narrower. The wavelengths of the vertical energetic ridge in the
spectrum of $u$ in figures \ref{fig:spec_1D}(a) and \ref{fig:spec_1D}(b) are approximately
$\lambda_x\approx3\delta_a$ and $\lambda_z\approx1.5\delta_a$. It is interesting to compare
this result with the aspect ratio of the vortex clusters \citep[3:1:1.5 in $x$, $y$ and $z$,
in][]{delAlamo06}, and of the sweep-ejection pairs \citep[4:1:1.5 in $x$, $y$ and $z$,
in][]{Lozano-Duran12}, although it is not immediately clear how $\delta_a$ should be related
to the wall-normal dimension of those threshold-based structures.

We finally compare the dynamical properties of the linear-stress layer with those of the
natural logarithmic layer. There is a central region within the active layer in which the
production and dissipation of the TKE match the ones of the natural logarithmic layer, but
the dissipation decays towards both ends of the active layer because the TKE is transported
away from it to compensate for the TKE deficit caused outside the active layer by the
elimination of the buffer and outer layer dynamics, instead of being dissipated in place. On
the other hand, the TKE production or, equivalently, the mean shear, compares well with the
natural logarithmic layer throughout the linear-stress layer when scaled with the mixing
length. The Corrsin parameter is approximately constant $2kS/\epsilon=8$, both in the active
layer and in the logarithmic layer, supporting the conclusion that the dynamics of the
eddies is dominated by the effect of the mean shear in both cases, and providing a rationale
for the use of the mixing length as a length scale for the structures.

\section{Conclusions}\label{sec:conc}

In conclusion, we demonstrate that the linear- {and constant-}stress layer of the present experiments
successfully reproduces the essential dynamics of the natural logarithmic layer, even in the
absence of a buffer and of an outer layer. Although there are some differences between the
two flows, such as a nonlinear mixing-length profile and the details of the TKE transport
and dissipation, the essential dynamics of the energy-producing and energy-containing
motions in the natural logarithmic layer is well-reproduced. Hence, the isolated system
introduced here should be useful to identify other intrinsic features of the logarithmic
layer {as well as the features that are not intrinsic to the logarithmic layer such as very large-scale structures}. In the present paper, we use it to support the {previous} idea that the logarithmic layer has
its own autonomous dynamics, which depend only weakly on inputs from other parts of the flow.
This is not to say that the other parts of the flow do not have an influence on the
logarithmic layer.
{In particular, we show that the dimensions of the longest structures depend on the flow above. Moreover, the size of self-similar logarithmic layer eddies is related to the height of the largest momentum-transferring eddies in the flow through the agency of the mean shear and momentum flux. However,} the sustenance of the logarithmic layer does not depend on {the other parts of the flow}. The
key problem in simulating an isolated logarithmic layer is how to limit the tendency of the
size of the turbulent structures to grow indefinitely as a result of the shear. To achieve
this objective, most of the previous attempts \citep{Flores10,Hwang15,Bae19} take a `minimal
box' approach, which controls the wall-normal eddy size by limiting the spanwise domain
dimension. However, the use of a minimal box inherently causes a significant portion of the TKE
to remain outside the range of resolvable scales, and their aggregate dynamics is projected
onto the streamwise- or spanwise-uniform modes. Instead, the present system represents
the opposite approach of creating a non-uniform shear profile by directly limiting the
wall-normal eddy size, which can accommodate the full scale dynamics of the
energy-containing eddies in the logarithmic layer.

\backsection[Funding]{This work was supported by the European Research Council under Coturb Grant No. ERC2014.AdG-669505.}

\backsection[Declaration of interests]{The authors report no conflict of interest.}

\appendix
\section{Effects of overdamping on the near-wall turbulence}\label{appA}

\begin{figure}
\centering
\includegraphics[width=0.5\linewidth]{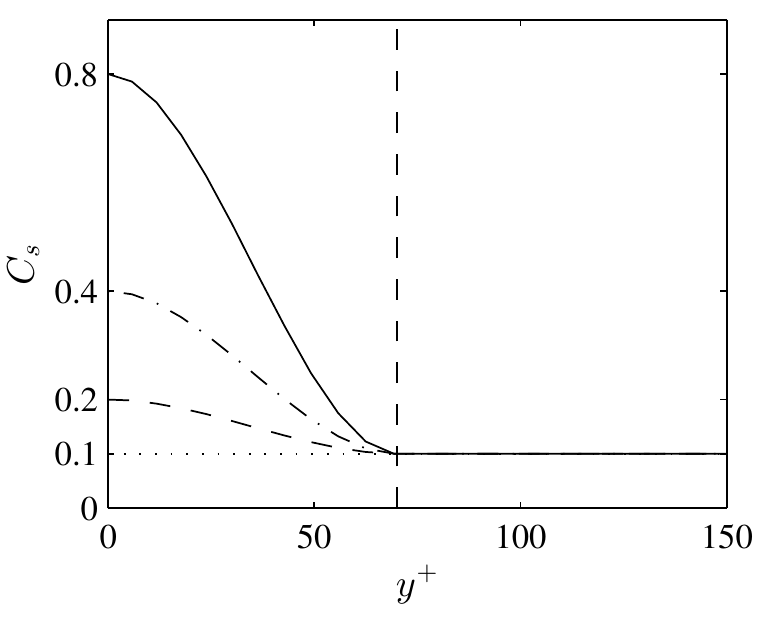}
\caption{The profiles of $C_s$ for $C_{s,w}=0.1$ (dotted), 0.2 (dashed), 0.4 (dash-dotted) and 0.8 (solid). The vertical line is at $y^+=70$.}
\label{fig:OD_Cs}
\end{figure}

As a preliminary trial, overdamping is applied in the buffer layer to test its effectiveness in suppressing the near-wall turbulence. Note that, because it was a test case, the simulation was conducted at a reduced spatial domain ($L_x/h=\pi$ and $L_z/h=\pi/2$) and spatial resolution ($\Delta x^+=\Delta z^+\simeq74$) compared to LB. All other simulation parameters are identical to LB. The overdamping is applied below $y^+=70$ and its degree is controlled by a parameter $C_{s,w}$, which represents the value of $C_s$ at the wall. The gradient of $C_s$ with respect to $y$ is set to be zero at the wall, and $C_s=0.1$ above $y^+=70$. For $0<y^+\leq70$, a cubic polynomial is fitted such that $C_s$ is continuous and differentiable at $y^+=70$. No van Driest damping is applied close to the wall. The profiles of $C_s$ for $C_{s,w}=0.1$, 0.2, 0.4 and 0.8 are shown in figure \ref{fig:OD_Cs}.

\begin{figure}
\centering
\includegraphics[width=0.75\linewidth]{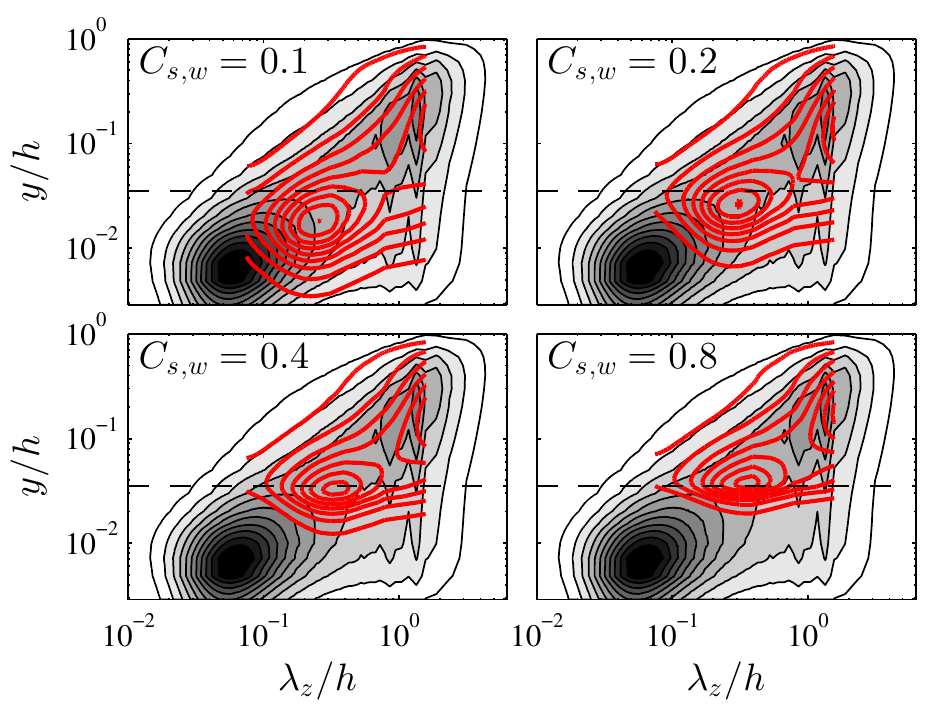}
\caption{Spanwise pre-multiplied spectra of $u$ for different values of $C_{s,w}$. The black (filled with gray scale colors in between) and red solid contours represent HJ06 and the overdamped experiments, respectively. The contour lines are drawn at multiples of $0.3U_\tau^2$. The horizontal dashed lines indicates $y^+=70$.}
\label{fig:OD_spec}
\end{figure}

As the most illustrative measure, the spanwise pre-multiplied spectra of $u$ for different values of $C_{s,w}$ are shown in figure \ref{fig:OD_spec}. Overdamping is effective at suppressing velocity fluctuations below $y^+=70$. However, with increasing $C_{s,w}$, the spectral signature of the near-wall cycle simply moves away from the wall and to the wider wavelengths instead of being eliminated at a fixed location. For the higher $C_{s,w}$ (especially for 0.8), it even protrudes into the non-overdamped region. This is in-line with the observation by \cite{Feldmann18} that the peak location of turbulent kinetic energy progressively moves outwards in the outer-scaled coordinates with increasing $C_s$, which they interpret as an effective reduction of the Reynolds number.

This problem does not exist when the buffer layer is suppressed by a modified body force. Figure \ref{fig:spec_1D}(b) shows that the spectral signature of the near-wall cycle is eliminated without leaving a residual in the cases LW and LN. Therefore, a modification of the body force is chosen as the preferred method for suppressing the buffer layer turbulence.

\bibliographystyle{jfm}
\bibliography{ref_masterfile.bib}

\end{document}